\documentclass[a4paper,11pt]{article}
\pdfoutput=1 

\usepackage{jheppub} 

\usepackage[T1]{fontenc} 
\usepackage{comment}

\usepackage[textsize=footnotesize,textwidth=2.cm]{todonotes}

\def\p{\partial}

\def\hu{{\hat u}}
\def\hv{{\hat v}}
\def\thv{ T_{\hat v}}
\def\thu{ T_{\hat u}}

\newcommand{\ba}{\begin{array}}
\newcommand{\ea}{\end{array}}
\newcommand{\bi}{\begin{itemize}}
\newcommand{\ei}{\end{itemize}}
\newcommand{\bea}{\begin{eqnarray}}
\newcommand{\eea}{\end{eqnarray}}
\newcommand{\be}{\begin{equation}}
\newcommand{\ee}{\end{equation}}

\title{\boldmath Modifications to Holographic Entanglement Entropy in Warped CFT}

\author[a]{Wei Song}
\author[a]{Qiang Wen}
\author[a]{Jianfei Xu}

\affiliation[a]{Yau Mathematical Sciences Center,Tsinghua University, Beijing, 100084, China}

\emailAdd{wsong@math.tsinghua.edu.cn}
\emailAdd{wenqqq@mail.tsinghua.edu.cn}
\emailAdd{jfxu@math.tsinghua.edu.cn}

\abstract{
In \cite{SWX} it was observed that asymptotic boundary conditions play an important role in the study of holographic entanglement beyond AdS/CFT. In particular, the Ryu-Takayanagi proposal must be modified for warped AdS$_3$ (WAdS$_3$) with Dirichlet boundary conditions.
In this paper, we consider AdS$_3$ and WAdS$_3$ with Dirichlet-Neumann boundary conditions. The conjectured holographic duals are warped conformal field theories (WCFTs), featuring a Virasoro-Kac-Moody algebra. We provide a holographic calculation of the entanglement entropy and R\'{e}nyi entropy using AdS$_3$/WCFT and WAdS$_3$/WCFT dualities. Our bulk results are consistent with the WCFT results derived by Castro-Hofman-Iqbal using the Rindler method. Comparing with \cite{SWX}, we explicitly show that the holographic entanglement entropy is indeed affected by boundary conditions.
Both results differ from the Ryu-Takayanagi proposal, indicating new relations between spacetime geometry and quantum entanglement for holographic dualities beyond AdS/CFT.

}

\begin{document}
\maketitle
\flushbottom

\section{Introduction}\label{intro}
The motivation of studying holography beyond AdS/CFT \cite{ADSCFT1} comes both from the bulk side and the boundary side.
In the bulk, to understand quantum gravity in the real world, it is necessary to consider spacetimes without being asymptotic to locally AdS. Progresses include dS/CFT correspondence \cite{dscft}, the Kerr/CFT correspondence \cite{KerrCFT1}, the WAdS/CFT \cite{warpedblackholes} or WAdS/WCFT \cite{DHH} correspondence, flat space holography \cite{flat1}, BMS group \cite{BMS}, etc.
On the boundary, there are richer physics beyond the critical point. Developments include Shr$\ddot{\text{o}}$dinger or Lifshitz spacetime/non-relativistic field theory duality \cite{lifshitz1}.

A three dimensional warped Anti-de Sitter space (WAdS$_3$), named and classified in \cite{warpedblackholes}, is a simple deformation of AdS$_3$, with $SL(2,R)\times U(1)$ isometry.  Due to a drastically different asymptotic behavior, a novel type of holographic dual, if exists, is expected. So far there are two versions of conjectures for the holographic duality, namely, the so-called WAdS$_3$/CFT \cite{warpedblackholes}, and the WAdS$_3$/WCFT correspondences \cite{DHH}.

The  WAdS$_3$/CFT$_2$ conjecture was motivated by the observation that the Bekenstein-Hawking entropy of WAdS$_3$ black holes can be captured by Cardy's formula \cite{warpedblackholes, Anninos08}. In support of the conjecture, a Dirichlet type of boundary conditions were found in \cite{CGR} under which the asymptotic symmetry group is generated by two Virasoro algebras.

The WAdS$_3$/WCFT conjecture is based on Dirichlet-Neumann boundary conditions \cite{Compere-Detournay}. The so-called two dimensional warped conformal field theory (WCFT) is defined by a warped conformal symmetry generated by a Virasoro-Kac-Moody algebra \cite{DHH}. WCFTs are usually not Lorentzian invariant. Much like the derivation of the Cardy formula \cite{Cardyformula} for a CFT, modular transformations lead to a Detournay-Hartman-Hofman (DHH) formula \cite{DHH} for the thermal entropy of WCFTs .  The agreement between the WAdS black hole entropy and the microscopic entropy calculated by the DHH formula gives evidence to this WAdS/WCFT holography.  See also \cite{Hofman-Strominger} for a pure field theoretic discussion of the symmetries, \cite{ChiralLiouville, Hofman:2014loa} for explicit examples, and \cite{Castro-Hofman-Sarosi} for partition functions of WCFTs. Furthermore, it was found in \cite{NBC} that Dirichlet-Neumann boundary conditions, later called the Comp\`{e}re-Song-Strominger (CSS) boundary conditions can also be imposed to AdS$_3$.  This leads to an alternative holographic dual to AdS$_3$, namely the AdS$_3$/WCFT correspondence. WAdS$_{3}$ can be embedded into string theory \cite{Guica-Strominger, Anninos08, ElShowk-Guica,Song-Strominger,Detournay-Guica}.
In some examples \cite{ElShowk-Guica, Song-Strominger, Detournay-Guica}, thermodynamics of WAdS$_3$ are even identical to those of AdS$_3$. Therefore  AdS$_3$/WCFT can be used as a good starting point of studying WAdS$_3$/WCFT.

So far  either  the framework of WAdS$_3$/CFT or WAdS$_3$/WCFT provides a microscopic explanation of the Bekenstein-Hawking entropy of WAdS$_3$ black holes. As a finer probe, entanglement entropy is expected to further test these two holographic dualities, which is the main focus of this paper.

In the context of AdS/CFT, the seminal work of Ryu and Takayanagi \cite{RT1,RT2} provides a powerful tool to understand the relationship between spacetime geometry and quantum entanglement. Ryu-Takayanagi proposal is now firmly established \cite{CHM,Faulkner,Hartman,LM} in the context of Einstein gravity, with a static,  asymptotically AdS spacetime in the bulk.
For WAdS$_3$, it is interesting to ask whether the Ryu-Takayanagi formula or the covariant  HRT formula \cite{HRT} is still valid, and how to derive/prove it if the answer is yes; or what is the analog if the answer is no.  The first attempt appeared in \cite{Anninos13}, where a tension was found between the HRT formula and the WAdS/CFT duality. See also \cite{Basanisi:2016hsh} for more discussions along this line.
Recently the holographic entanglement entropy for WAdS$_3$ spacetime with Dirichlet boundary conditions was calculated in \cite{SWX} with a modified Lewkowycz-Maldacena \cite{LM} prescription. The result is consistent with the WAdS/CFT  duality, but is different from a direct use of the HRT formula. A key observation is that asymptotic boundary conditions play an important role in such holographic dualities beyond AdS/CFT. In this paper we consider  AdS and WAdS with the Dirichlet-Neumann boundary conditions.

We study the holographic entanglement entropy in the context of AdS/WCFT and WAdS/WCFT correspondences.  The approach we take is the Rindler method \cite{CHM}, which was developed in the context of AdS/CFT. On the CFT side, a certain conformal transformation maps  an entanglement entropy to the thermal entropy of a Rindler or hyperbolic space. Using the dictionary of AdS/CFT, the bulk counterpart of this procedure is to perform a coordinate transformation, mapping a certain region of Poincar\'{e} AdS to a hyperbolic black hole. The Bekenstein-Hawking entropy of the hyperbolic black hole then calculates the entanglement entropy holographically.
Generalization  to WCFT was carried out in \cite{CHI}, where the role of conformal transformations is now played by warped conformal transformations allowed by the symmetry of WCFT.
The current paper provides a  holographic calculation on the gravity side\footnote{In \cite{CHI}, a bulk calculation was carried out for lower spin gravity \cite{Hofman:2014loa}, which does not have an usual geometric description.}.
Using the strategy of doing quotient on WAdS spacetime, we find that the analog of hyperbolic black holes are the WAdS black strings. We proposed that the thermal entropy of the WAdS black string gives the holographic calculation of entanglement entropy of WCFT. Explicitly agreement with the field theory result is found using the dictionary of AdS/WCFT and WAdS/WCFT.
R\'{e}yni entropy in the bulk AdS$_3$ is also calculated and showed to be consistent with the WCFT calculation.

To summarise, we provide a holographic calculation of entanglement entropy and R\'{e}yni entropy for AdS and WAdS under the Dirichlet-Neumann boundary conditions. Our bulk results agree with those of WCFT \cite{CHI}, further supporting the conjectured AdS/WCFT and WAdS/WCFT dualities.
Holographic entanglement entropy for WAdS$_3$ in this paper differs from that of \cite{SWX}, the reason of which is the different choices of asymptotic boundary conditions.
Both this paper and \cite{SWX} differ from the Ryu-Takayanagi proposal, indicating new relations between spacetime geometry and quantum entanglement for holographic dualities beyond AdS/CFT.

The layout of this paper is the following. In section \ref{secblackstring}, we briefly review WCFT, AdS$_3$/WCFT and WAdS$_3$/WCFT. In section  \ref{secfeildtheorystrory} we re-calculate the entanglement entropy for WCFT, along the lines of \cite{CHI}.  In  section  \ref{secgravitystory}, a bulk calculation is proposed.  Then we match the results calculated on both sides in section \ref{secmatching}. In section \ref{renyi} we calculate the Reyni entropy, and show the agreement between the bulk and boundary calculations.

\section{ AdS$_3$/WCFT and WAdS$_3$/WCFT }\label{secblackstring}
In this section, we briefly review AdS$_3$/WCFT and WAdS$_3$/WCFT, and set up notations and conventions.
In section \ref{wcft} we list a few properties of WCFT.  Section \ref{2.2} is for AdS$_3$/WCFT, and section \ref{2.3} is  for WAdS$_3$/WCFT.

\subsection{WCFT}\label{wcft}
In this subsection, we discuss warped conformal field theory (WCFT) with a pure field theoretical setup.
In \cite{Hofman-Strominger} it was shown that a two dimensional local field theory with translational invariance $z=z'+z_0,\quad w=w'+w_0$ and a chiral scaling symmetry $z=\gamma z'$ will have some enhanced symmetries. One minimal option is
to have the following local symmetry
\bea\label{warpedmapping} z= f(z'),\qquad w=w'+g(z')\,.\eea
The above property was later used as a definition of WCFT in \cite{DHH}.
On a cylinder, the conserved charges can be written in terms of Fourier modes. The WCFT algebra is \cite{DHH}
\begin{align}\label{wcftalg}
[L_n,L_m]=&(n-m)L_{n+m}+\frac{c}{12}(n^3-n)\delta_{n+m}\,,
\cr
[L_n,P_m]=&-m P_{n+m}\,,
\cr
[P_n,P_m]=&\frac{k}{2}n\delta_{n+m}\,,
\end{align}
which describes a Virasoro algebra with central charge $c$ and Kac-Moody algebra with level $k$, and furthermore the Kac-Moody generators transform canonically under the action of Virasoro generator. We will hereafter refer to (\ref{wcftalg}) as the canonical WCFT algebra.

So far there are some concrete examples of WCFT. The chiral Liouville gravity \cite{ChiralLiouville} is a bosonic model, and  can be obtained from Chern-Simons formulation of Einstein gravity with CSS boundary conditions, analogous to the usual Liouville theory under Brown-Henneaux boundary conditions. The Fermionic models discussed in \cite{Hofman:2014loa}  are closely related to  holography for the so-called lower spin gravity.  Partition functions of the Fermionic models were calculated in \cite{Castro-Hofman-Sarosi}.

\subsection{AdS$_3$/WCFT}\label{2.2}
In this subsection we lay out a few properties of the AdS$_3$/WCFT correspondence, focusing on black holes and black strings. We start with the algebra of asymptotic symmetry  under the CSS boundary conditions in section \ref{adswcft}, and show how to relate it to canonical Virasoro-Kac-Moody algebra defining a WCFT and how the black hole entropy can be reproduced using the DHH formula in section \ref{2.2.2}. In section \ref{2.2.3}, we discuss black strings and show how to effectively calculate the thermodynamic quantities using a black hole.

\subsubsection{AdS$_3$/WCFT$_{(\hu,\hv)}$}\label{adswcft}
In the Fefferman-Graham gauge, solutions to three-dimensional Einstein gravity with a negative cosmological constant can be written as
\be
{ds^2\over\ell^2}={d\eta^2\over \eta^2}+\eta^2\Big(g^{(0)}_{ab}+{1\over \eta^2}g^{(2)}_{ab}+{1\over \eta^4}g^{(4)}_{ab}\Big)dx^a dx^b\,,
\ee
where $\eta$ is the radial direction, and $x^a,\, a=1,2$ parametrize the boundary.
The Dirichlet boundary conditions can be written as
\bea\label{dirichlet}
\delta g_{ab}^{(0)}=0\label{dirichlet}\,,
\eea
under which the asymptotic symmetry are generated by two copies of Virasoro algebra. This indicates that Einstein gravity on asymptotically AdS$_3$ spacetime under the Brown-Henneaux boundary conditions is holographically dual to a two dimensional conformal field theory.
The original Brown-Henneaux boundary conditions further specifies that
\bea
&&\phi\sim\phi+2\pi\label{bsc}\,.
\eea
This corresponds to put the CFT on a cylinder.
More generally, we can consider other Dirichlet boundary conditions with (\ref{dirichlet}) but a different choice of (\ref{bsc}).
On the boundary, this corresponds to put the CFT on a different manifold $\mathcal{N}$ with a non-dynamical metric $g_{ab}^{(0)}$,
and two different choices are related by a conformal transformation.
In the bulk, this corresponds to a different foliation of AdS$_3$,  locally related by coordinate transformation.

In \cite{NBC}, Dirichlet-Neumann boundary conditions is considered for AdS$_3$ in Einstein gravity \bea
\delta g^{(0)}_{\pm-}&=&0\,,  \,  \p_- g^{(0)}_{++}=0\,,\label{css1}\\
\delta g^{(2)}_{--}&=&0\,. \label{css}
\eea The asymptotic symmetry for AdS$_3$ is generated by a Virasoro-Kac-Moody algebra, indicating a WCFT as the holographic dual.
In particular, under the choice \bea g^{(0)}_{\hu\hv}=1, \quad g^{(0)}_{\hv\hv}&=&0 , \quad \p_{\hv}g^{(0)}_{\hu\hu}=0,\quad g^{(2)}_{\hv\hv}=T^{2}_{\hv} \label{tv}\label{uhat1}\\
(\hu,\,\hv)&\sim& (\hu+2\pi,\, \hv+2\pi)\label{uhat}
\eea
The WCFT is put on manifold with a fixed spatial circle (\ref{uhat}).
 The Virasoro-Kac-Moody algebra can be written as\footnote{We notice that \cite{NBC} misses the anomalous term in the commutator between $\tilde{L}_n$ and $\tilde{P}_m$. }
\begin{align}\label{tildealgebra}
[\tilde{L}_n,\tilde{L}_m]=&(n-m)\tilde{L}_{n+m}+\frac{\tilde{c}}{12}(n^3-n)\delta_{n+m}\,,
\cr
[\tilde{L}_n,\tilde{P}_m]=&-m\tilde{P}_{m+n}+m\tilde{P}_0\delta_{n+m}\,,
\cr
[\tilde{P}_n,\tilde{P}_m]=&\frac{\tilde{k}}{2}n\delta_{m+n}\,,
\end{align}
with  the central charge and Kac-Moody level \begin{align}
\tilde{c}=&\frac{3\ell}{2G}\,, \qquad \tilde{k}=-\frac{\ell\thv^2}{ G}\,.
\end{align}
Similarly, different choice of the boundary manifold will be related by warped conformal transformations. In this paper,  Comp\`{e}re-Song-Strominger (CSS) boundary conditions refer to consistent boundary conditions (\ref{css1}), (\ref{css}) with any fixed boundary manifold.
See also \cite{Troessaert:2013fma} for other choices of consistent boundary conditions.

Consider BTZ black holes
\bea\label{btz}
ds^2&=&\ell^2 \left(\thu^2d\hu^2+2 r d\hu d\hv+\thv^2 d\hv^2+\frac{dr^2}{4 \left(r^2-\thu^2 \thv^2\right)}\right)\,,\\
\label{btzspatial}(\hu,\,\hv)&\sim& (\hu+2\pi,\, \hv+2\pi)
\,.\eea
The local isometry is $SL(2,R)\times SL(2,R)$,
while only the $U(1)\times U(1)$ part is globally well-defined due to the spatial circle (\ref{btzspatial}).
The Bekenstein-Hawking entropy is
\be S_{BH}=\frac{\pi \ell}{2G}\left(\thu+\thv\right)\,.\ee
The phase space under the conditions (\ref{uhat1})-(\ref{uhat}) contains  the BTZ black holes with fixed $T_{\hv}$ but arbitrary $T_{\hu}$ and all their Virasoro-Kac-Moody desendents. We will call this phase space $\mathcal{H}_{{\tilde P}_0}$.
The nonzero conserved charges $\tilde{P}_0$ and $\tilde{L}_0$ that are associated with Killing vectors $\partial/\partial_{\hat{v}}$ and $\partial/\partial_{\hat{u}}$ respectively for the BTZ metric (\ref{btz}) can be calculated,
\be \tilde{P}_0=Q[\partial_{\hat{v}}]=-\frac{\ell\thv^2}{4 G}\,, \qquad\tilde{L}_0=Q[\partial_{\hat{u}}]=\frac{\ell \thu^2}{4G}\,.
\ee
As was argued in \cite{DHH}, the thermal entropy of a theory with the symmetry (\ref{tildealgebra}) can be written as a Cardy-like formula
\begin{align}\label{thermalentropyofblackstring}
S_{micro}=2\pi\sqrt{-\tilde{P}_0^{vac}\tilde{P}_0}+2\pi\sqrt{-\tilde{L}_0^{vac}\tilde{L}_0}\,,
\end{align}
where $\tilde{P}_0^{vac}$ and $\tilde{L}_0^{vac}$ are the vacuum values of the zero-mode charges.
A natural way to find the vacuum charges is to rewrite the metric of BTZ black holes  (\ref{btz}) in the Schwarzschild  form, where it is easy to find that Global AdS$_3$ is the vacuum, with
\begin{align}\label{vacuumtemperatures} \thu^{vac}=\pm \frac{i}{2},\qquad \thv^{vac}=\pm\frac{i}{2}\,,
\end{align}
or equivalently,
\begin{align}\label{tPvac}
\tilde{P}_0^{vac}=-\tilde{L}_0^{vac}=\frac{\tilde{c}}{24}\,.
\end{align}
Plugging these vacuum values into (\ref{thermalentropyofblackstring}) the microscopic formula (\ref{thermalentropyofblackstring}) matches the Bekenstein-Hawking entropy of the BTZ black holes (\ref{btz})
\begin{align}
S_{micro}=\frac{ \pi \tilde{c}}{3}\left( \thu+\thv\right)=S_{BH}\,.
\end{align}

\subsubsection{From WCFT$_{(\hu,\hv)}$ to WCFT$_{(\hat x,\hat t)}$ }\label{2.2.2}
Note that the algebra (\ref{tildealgebra}) is different from the canonical WCFT algebra (\ref{wcftalg}). To make distinctions, hereafter we will use the coordinates as subscripts. We denote the geometry (\ref{btz}) by AdS$_{(\hat{u},\hat{v},r)}$, and denote the field theory defined by the algebra (\ref{tildealgebra}) by WCFT$_{(\hat{u},\hat{v})}$.
The algebra (\ref{tildealgebra}) and the canonical WCFT algebra (\ref{wcftalg}) are related by a charge redefinition as was shown in \cite{DHH}
\begin{align}\label{ensemblechange}
\tilde{P}_n=\frac{2P_0 P_n}{k}-\frac{P_0^2 \delta_n}{k}\,, \qquad \tilde{L}_n=L_n-\frac{2P_0 P_n}{k}+\frac{P_0^2 \delta_n}{k}\,.
\end{align}
with $c=\tilde{c}$.
Note that none of the classical solutions (\ref{btz}) has a higher Kac-Moody hair, or in other words, they all satisfy  $P_{n\neq0}=0$.
For states with $P_{n\neq0}=0$, (\ref{ensemblechange}) amounts to a coordinate transformation,
\begin{align}\label{nlct}
\hu=\hat{x}\,, \qquad \hv=\frac{k \hat{t}}{2P_0}+\hat{x}\,.
\end{align}
where $k$ is just the level of the Kac-Moody algebra in (\ref{wcftalg}). The mappings between zero modes are
\begin{align}\label{P0L0}
P_0=-\sqrt{\tilde{P}_0 k}=-\frac{T_{\hat{v}}}{2}\sqrt{-\frac{\ell k}{G}}\,,\qquad L_0=\tilde{ L}_0+\tilde{ P}_0=\frac{\ell}{4G}(\thu^2-\thv^2)\,.
\end{align}
Throughout this paper we will consider a negative $k$. This will make $P_0$ real for all BTZ black holes, and pure imaginary for global AdS$_3$. Given $k$, (\ref{ensemblechange}) maps the phase space with fixed $\tilde{P}_0$( or equivalently, fixed $\tilde{k}$ or fixed $T_\hv$) to a fixed $P_0$ sector of the phase space of WCFT$_{(\hat x,\hat t)}$.
The union of the phase spaces with all different $\tilde{P}_0$, namely $\cup \mathcal{H}_{\tilde{P}_0}$, will be mapped to the entire phase space of WCFT$_{(\hat x,\hat t)}$ with level $k$.
As discussed before,  $\mathcal{H}_{\tilde{P}_0}$ contains all BTZ black holes with fixed $T_\hv$ and all their Virasoro-Kac-Moody descendants. By allowing $P_0$ to be pure imaginary at a single point( the vacuum), and real elsewhere, the phase space of  WCFT$_{(\hat x,\hat t)}$ contain global AdS$_3$ and all BTZ black holes and their descendants. Note that $P_0^2<0$ for the global AdS$_3$ is due to the choice $k<0$ which is more convenient for the black hole sector. Alternatively, if we take $k>0$, $P_0^{vac}$ is also real. Therefore the theory is unitary at least in the vacuum sector. There might be some instabilities in the black hole sectors, which we will not going to discuss in the present paper. More discussions about the spectrum and representations can be found in \cite{DHH,NBC}.

The spatial circle (\ref{btzspatial}) leads to a spatial circle in the ${\hat x},\,{\hat t}$ coordinates,  \be spatial : \quad ({\hat x},{\hat t})\sim ({\hat x}+2\pi, {\hat t})\,.\ee
The temperature of the black hole is translated to a thermal circle of WCFT$_{(\hat{x},\hat{t})}$
\begin{align}\label{thermalcirclehxht}
thermal : \quad ({\hat x},{\hat t})\sim \left({\hat x}+i\frac{\pi}{T_{\hat{u}}}, {\hat t}-i\pi\sqrt{\frac{-\ell }{k G}}\left(\frac{T_{\hat{v}}}{T_{\hat{u}}}+1\right)\right)\,.
\end{align}
The vacuum expectation values for the zero modes (\ref{tPvac}) are then \be\label{P0L0vac} P_0^{vac}=\frac{i}{4}\sqrt{-\frac{\ell k}{G}},\quad L_0^{vac}=0\,.\ee
Plugging the above into DHH formula \cite{DHH} in this ensemble,
{ \begin{align}
S_{DHH}=& -\frac{4\pi i P_0P_0^{vac}}{k}+4\pi\sqrt{-\left(L_0^{vac}-\frac{(P_0^{vac})^2}{k}\right)\left(L_0-\frac{P_0^2}{k}\right)}
\cr
=&\frac{ \pi c}{3}\left( \thu+\thv\right)\,,
\end{align}}
we again reproduce the macroscopic entropy of the black hole, $S_{DHH}=S_{BH}$.
Note that the choice of $k$ will affect some details of the dictionary between the bulk and the WCFT$_{(\hat x,\hat t)}$. For example,  the vacuum charge $P_0^{vac}$ is  $k-$dependent.  But the combination ${(P_0^{vac})^2\over k}$ and the entropy are both insensitive to the choice of $k$.

\subsubsection{From WCFT$_{(u,v)}$ to WCFT$_{( \hu,\hv)}$}\label{2.2.3}

If we uncompactify the spatial circle (\ref{btzspatial}), we get a BTZ black string with an infinite horizon
\be ds^2=\ell^2 \left(T_u^2du^2+2 r du dv+T_v^2 dv^2+\frac{dr^2}{4 \left(r^2-T_u^2 T_v^2\right)}\right)\,.\label{btzstring}\ee
Now we consider the black string (\ref{btzstring}) on an arbitrary spatial interval
\be \{ (u,\,v)|\,u={\Delta u}\left(-{1\over2}+\tau\right) ,\quad v={\Delta v}\left(-{1\over2}+\tau\right),\quad \tau\in[0,1]\}\label {uvspatial}\,.\ee
If the interval is very large, for example if $\Delta u\rightarrow\infty$, the system on (\ref{uvspatial}) is equivalent to
a black hole with the periodicity
\be (u,\,v)\sim (u+\Delta u,\,v+\Delta v)\,. \label{uvbh}\ee
In particular, the total charges, and the thermal entropy for BTZ black string on the interval (\ref{uvspatial}) are always the same as the BTZ black hole with a spatial circle (\ref{btzspatial}).
As we will see later, we will always encounter the systems with $\Delta u\rightarrow \infty$. Hereafter, we will often view theories on a large spatial interval as on a spatial circle without further explanations.
The total charges and entropy of the black hole (\ref{uvbh})/black string (\ref{uvspatial}) are the same as
those of (\ref{btz}) (\ref{btzspatial}) with the mapping
\be u=\frac{\Delta u}{2\pi} {\hat u},\quad v=\frac{\Delta v}{2\pi} {\hat v},\quad T_u=2\pi{\thu\over\Delta u} ,\quad T_v=2\pi{\thv\over\Delta v},\quad r=\frac{4\pi^2}{\Delta u\Delta v}\hat{r}, \ee Thus the thermal entropy is given by
\begin{align}
S=\frac{\pi\ell}{2G}\left({\hat T}_u+{\hat T}_v\right)=\frac{\ell}{4G}\left(\Delta u T_u+\Delta v T_v\right) \,.\label{suv}
\end{align}
As discussed before, the CSS boundary conditions (\ref{tv}) can be imposed with an arbitrary boundary manifold. Correspondingly,  the dual field theory from the asymptotic symmetry analysis with a spatial circle (\ref{uvbh}) are denoted by WCFT$_{(u,v)}$. Furthermore, by a charge redefinition, we can also get WCFT$_{( x, t)}.$

\subsection{WAdS$_3$/WCFT}\label{2.3}

Warped AdS$_3$ (WAdS$_3$) appears in many context, including three dimensional theories \cite{TMG1,TMG2, NMG}, and some six dimensional theories \cite{Anninos08}, \cite{ElShowk-Guica, Song-Strominger}, \cite{Bena-Guica-Song, Azeyanagi-Hofman-Song-Strominger}, \cite{Detournay-Guica}.

In this paper, we consider a class of locally WAdS$_3$ spacetimes, the warped black string solutions from  consistent truncations of IIB string theory \cite{Detournay-Guica}.
The metric can be written as
\bea\label{blackstring}
ds^2&=&\ell^2 \left(T_u^2\left(1+\lambda ^2 T_v^2\right)-\lambda^2 r^2) du^2+2 r du dv+T_v^2 dv^2+\frac{(1+\lambda^2 T_v^2)dr^2}{4 \left(r^2-T_u^2 T_v^2\right)}\right)\,.
\eea
The parameter $\lambda$ is the warping parameter, whose existence breaks the $SL(2,R)\times SL(2,R)$ local isometry of AdS$_3$ to the $SL(2,R)\times U(1)$ local isometry of WAdS$_3$. When $\lambda=0$, the warped black string metric goes back to the BTZ black string (\ref{btzstring}).
A key feature of these models is that the thermodynamic properties of these warped black strings are independent of the warping factor $\lambda$.
When $\lambda\neq0,$ we have to keep the spatial circle of (\ref{blackstring}) uncompactified, otherwise there will be closed time-like curves. However, similar to BTZ black string, for the purposes of discussing the thermodynamics, calculations on an infinitely large spatial interval
\be
 \{(u,\,v)|\, u={\Delta u}\left(-{1\over2}+\tau\right) ,\quad v={\Delta v}\left(-{1\over2}+\tau\right),\quad \tau\in[0,1]\}\,,
 \ee
 is effectively the same as on a spatial circle
 \be\label{spatialcircleuvr}
  (u,v)\sim (u+\Delta u,v+\Delta v)\,.
 \ee
With the Dirichlet-Neumann type of boundary conditions \cite{Detournay-Guica,Compere-Detournay,CGR}, the asymptotic symmetry of (\ref{blackstring}) is generated by a chiral stress tensor and a U(1) current. On a cylinder, the symmetry is generated by the non-canonical Virasoro-Kac-Moody algebra (\ref{tildealgebra}). The thermal circle of (\ref{blackstring}) is given by
\be
thermal\, circle: \quad (u,v) \sim \left(u+{\pi i\over T_u},~ v-{\pi i\over T_v}\right)\,, \label{tcuv}
\ee
 All the discussions in the previous subsection for AdS/WCFT can be repeated here in the context of WAdS/WCFT.

\section{Entanglement entropy for WCFT}\label{secfeildtheorystrory}
In this section we compute the entanglement entropy of a single interval in WCFT with an adapted version of the Rindler method \cite{CHM}, following the main steps of \cite{CHI}. Note that this section is a purely field theoretical calculation, without any reference to holography.

Based on \cite{CHMbase1,CHMbase2,CHMbase3,CHMbase4}, \cite{CHM} developed the Rindler method to derive the Ryu-Takayanagi formula for spherical entangling surfaces in the context of AdS$_{d+1}$/CFT$_d$. It contains both a field theory story and a gravity story. On the field theory side,
a spherical entangling surface $S^{d-2}$ at constant time slice is considered in the vacuum state {\footnote{ In particular, when $d=2$ this method also works at finite temperature. In Appendix \ref{A2.1} we also explicitly extend this method to a covariant version in three dimensions and reproduce the HRT formula. }} of a $d$-dimensional CFT on $\mathbb{R}^d$.
A certain conformal transformation maps the causal development of the subsystem to a Rindler space or a hyperbolic space, and maps the reduced density matrix of the former to a thermal density matrix of the later.  Therefore the entanglement entropy is mapped to the thermal entropy of the Rindler/hyperbolic space. We will call such a transformation a {\it Rindler transformation}. The gravity side story is just the bulk extension of the field theory side story through the AdS/CFT dictionary.  On the gravity side, a transformation which maps Poincar\'{e} AdS$_{d+1}$ space to a hyperbolic black hole is needed.  The thermal entropy of the CFT on the hyperbolic space is therefore the Bekenstein-Hawking entropy of the hyperbolic black hole.

The field theory side story of the Rindler method \cite{CHM} was extended to WCFT in \cite{CHI}, which considers arbitrary single intervals and WCFTs with an arbitrary thermal or spatial circle. Moreover, \cite{CHI} calculate the thermal entropy of the WCFT on the ``Rindler space''. Before we go through the story explicitly, we would like to emphasize three main points for the extension.
\begin{itemize}
\item Firstly, as WCFT is not Lorentzian invariant, the entanglement entropy of an arbitrary interval would rely on both the length and direction of the interval. Also, as we will see later, the subregion region covered by the ``Rindler spacetime'' is a stripe, instead of a diamond shape.

\item Secondly, a {\it Rindler transformation} should be a symmetry of the field theory to implement a unitary transformation   between the reduced density matrix and the thermal density matrix. So the entanglement entropy equals to the thermal entropy of the ``Rindler'' space. For WCFTs,  a {\it Rindler transformation}  should be a warped conformal transformation in the form of (\ref{warpedmapping}) rather than a conformal transformation.
\item Thirdly, it is possible to calculate the thermal entropy of a WCFT on the ``Rindler'' space under some approximations. The key point is to consider the ``Rindler" space as an infinitely large spatial circle. Effectively, we put the theory on a torus. Then we can use the DHH formula (the warped Cardy formula) to calculate the thermal entropy.
Following the logic of deriving the Cardy formula in a CFT \cite{Cardyformula}, Detournay, Hartman and Hofman calculated the thermal entropy for WCFT by using properties of modular transformations \cite{DHH}. The idea is to find a warped conformal transformation which exchanges the thermal circle and the spatial circle, and consequently the asymptotic density of states can be written in terms of vacuum expectation values of the energy and angular momentum. In CFT, this is the $S$-transformation. We will use this terminology for WCFT as well.
Putting all together, \cite{CHI} managed to write down  a formula of entanglement entropy of an arbitrary interval  in WCFT.
 Note that WCFT is not modular invariant, we need to keep track of the anomalies.
\end{itemize}

To recapitulate, two types of warped conformal mappings are essential. The {\it Rindler transformation}
 is used to map entanglement entropy to thermal entropy, while the $S$-transformation is used to estimate the thermal entropy.

In this paper, we use a more general set of {\it Rindler transformations}, with one additional parameter $\alpha$ as compared to \cite{CHI}. On the WCFT side, we recover the results of \cite{CHI} if we choose $\alpha=0.$ However, as will be seen later, to match the gravity results, we have to choose $\alpha=\sqrt{\frac{-\ell}{Gk}}\pi$ instead.

\begin{figure}

\centering
\includegraphics[width=0.7 \textwidth]{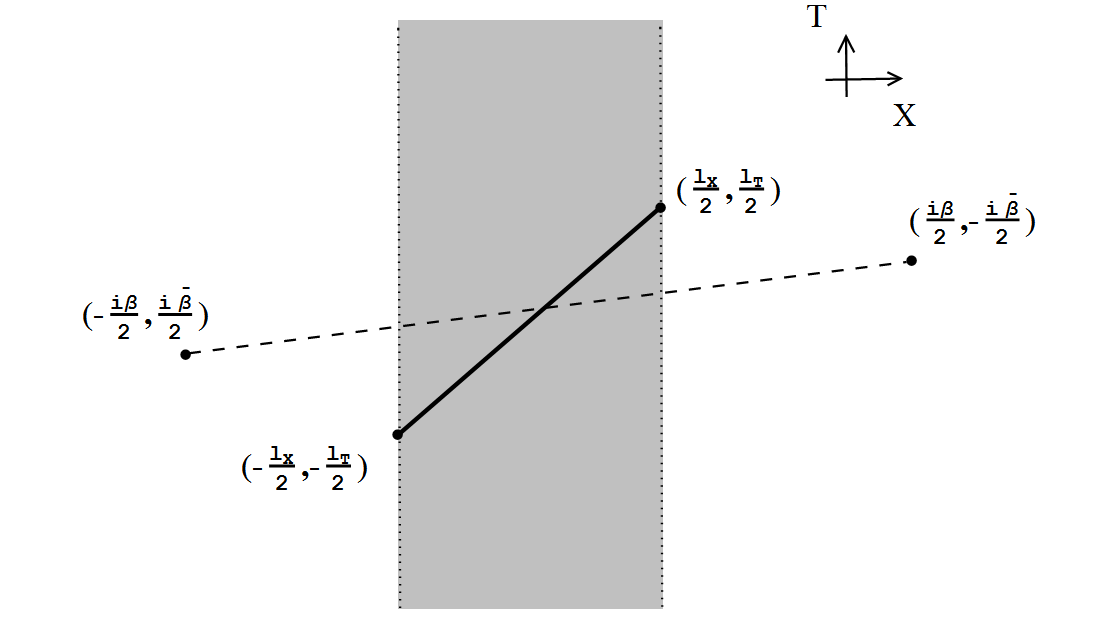}

\caption{\label{Fig2} Diagram that depicts the region of (X,T) covered by the (x,t) space, which is the shaded strip. The solid line segment is the interval (\ref{intervalXT}), and the identification of the dashed line gives the thermal circle. }
\end{figure}

We begin with the WCFT$_{(X,T)}$, and consider the following interval
\begin{align}\label{intervalXT}
\mathcal{A}:~~~~\{ (X,\,T)|\,X={l_X}\left(-{1\over2}+\tau\right) ,\quad T={l_T}\left(-{1\over2}+\tau\right),\quad \tau\in[0,1]\}\,.
\end{align}
To calculate the entanglement entropy of this interval, the key is to find a suitable warped conformal mapping to a plane with a thermal identification, i.e. the analog of Rindler/Hyperbolic space. Here we would like to apply the following warped conformal mapping which satisfies (\ref{warpedmapping})
\bea\label{warpedmapping1}
&&\frac{\tanh\frac{\pi X}{\beta}}{\tanh\frac{\textit{l}_X \pi}{2 \beta}}=\tanh\frac{\pi x}{\kappa}\,,
\qquad\\
&&T+\left(\frac{\bar{\beta}}{\beta}-\frac{\alpha }{\beta}\right)X=t+\left(\frac{\bar{\kappa}}{\kappa}-\frac{\alpha }{\kappa}\right)x\equiv s\,.
\label{spectralflow}\eea
Let us denote the $(x,t)$ space by $\mathcal{H}$.
This warped conformal mapping (\ref{warpedmapping1})-(\ref{spectralflow}) have the following key features:
\begin{enumerate}
\item
 $\mathcal{H}$ covers a strip region with $-\frac{\textit{l}_X}{2}<X<\frac{\textit{l}_X}{2}$, i.e. the shaded region in Fig. \ref{Fig2}.
 The mapping induces a thermal circle in $\mathcal{H}$:
\begin{align}\label{thermalcircle}
thermal:~~~~(x,t)\sim (x+i \kappa, t-i\bar{\kappa})\,.
\end{align}
\item The mapping  (\ref{warpedmapping1})-(\ref{spectralflow}) takes the form of (\ref{warpedmapping}), and hence is a symmetry transformation of the field theory. This indicates the field theory on the ``Rindler space'' is also a warped CFT, which we denote as WCFT$_{(x,t)}$. As was argued in \cite{CHM},  the entanglement entropy of the interval $\mathcal{A}$ equals to the thermal entropy of the ``Rindler space''  $\mathcal{H}$,
\begin{align}
S_{EE}=-\text{tr}\left(\rho_{\mathcal{A}}\log\rho_{\mathcal{A}}\right)=S_{thermal}(\mathcal{H})\,.
\end{align}
\item More precisely, the mapping (\ref{warpedmapping1}) is a confromal transformation between $x$ and $X$, while the mapping (\ref{spectralflow}) is a spectral flow. Define a new variable $s$, the spectral flow parameter  is $\frac{\bar{\beta}}{\beta}-\frac{\alpha }{\beta}$ for the flow from $(X,T)$ to $(X,s)$, and $\frac{\bar{\kappa}}{\kappa}-\frac{\alpha }{\kappa}$ for the flow from $(x,t)$ to $(x,s)$. \item The combination of $t$ and $x$ in (\ref{spectralflow}), denoted by $s$, is invariant under Galileo boost $t\rightarrow t+vx$. Going around the thermal circle (\ref{thermalcircle}) maps $s$ to $s-i\alpha$. When $\alpha=0$, $s$ is invariant under thermal identifications (\ref{thermalcircle}).
 As we will see more explicitly later, $\alpha$ will also affect how we take the cylinder limit of a torus.
\end{enumerate}


The divergence of $S_{EE}$, which arises from the short distance entanglement near the end points of the interval $\mathcal{A}$ is now mapped to the divergence of the size of the thermal system $\mathcal{H}$. To see this explicitly we introduce a cutoff $\epsilon$ and define a regularized interval
\begin{align}\label{regulatedinterval}
\mathcal{A}:~~~~\{ (X,\,T)|\,X={(l_X-2\epsilon)}\left(-{1\over2}+\tau\right) ,\quad T={\left(l_T-2 \frac{\textit{l}_T}{\textit{l}_X}\epsilon\right)}\left(-{1\over2}+\tau\right),\quad \tau\in[0,1]\}\,.
\end{align}
Notice that the factor in front of the cutoff in the $T$ direction is chosen to guarantee that the regularized interval is contained in the original interval. Using the mapping (\ref{warpedmapping1}), the image of the regularized interval in $(x,t)$ coordinates is given by \begin{align}
~~~~&\{ (x,\,t)|\,x={2\pi a}\left(-{1\over2}+\tau\right) ,\quad T={2\pi \bar{a}}\left({1\over2}-\tau\right),\quad \tau\in[0,1]\}\,,
\\
\quad &2\pi a =\frac{\kappa\zeta}{\pi}, \quad 2\pi\bar{a}=\frac{\bar{\kappa}-\alpha}{\pi}\zeta-\textit{l}_T-\textit{l}_X\left(\frac{\bar{\beta}-\alpha}{\beta}\right)\,,\label{abara}
\end{align}
where
\begin{align}
\zeta=\log\left(\frac{\beta}{\pi\epsilon}\sinh\frac{\textit{l}_X \pi}{\beta}\right)+\mathcal{O}(\epsilon)\,.
\end{align}
Here we have taken the small $\epsilon$ limit and only kept the leading term in the expansion. Under this limit, $\zeta$ becomes very large.

The image of the regularized interval in the $(x,t)$ coordinates is of infinite length, hence we expect the edge effects can be omitted.
Therefore we can identify its endpoints, and denote this identification as the spatial circle
\begin{align}\label{spatialcircle}
spatial \,circle: (x,t)\sim (x+2\pi a,t-2\pi \bar{a})\,.
\end{align}
This circle together with the thermal circle (\ref{thermalcircle}) form a torus, and the partition function calculated on this torus can be denoted as
$Z_{\bar{a}|a}(\bar{\kappa}|\kappa)$.
It is useful to perform a further warped conformal mapping
\begin{align}\label{ctcanonical}
\hat{x}=\frac{x}{a},\qquad \hat{t}=t+\frac{\bar{a}}{a}x\,,
\end{align}
which changes an arbitrary torus to a canonical torus with a canonical (spatial) circle with $(a,\bar{a})=(1,0)$, and a thermal circle independent of the parameters $\kappa,\,\bar \kappa$:
\begin{align}
(\hat{x},\hat{t})\sim (\hat{x}+2\pi,\hat{t})\sim (\hat{x}+i\hat{\kappa},\hat{t}-i\bar{\hat{\kappa}})\,,
\end{align}
with
\begin{align}\label{hkappa1}
\hat{\kappa}=\frac{\kappa}{a}=\frac{2\pi^2}{\zeta}\,,\qquad \hat{\bar{\kappa}}=\bar{\kappa}-\frac{\bar{a}}{a}\kappa=\alpha+\pi\frac{l_{T}+\frac{\bar{\beta}-\alpha}{\beta} l_{X}}{\zeta}\,.
\end{align}
To calculate the partition function, one needs to perform a modular transformation $S$, which exchanges the spatial and thermal circles.
The partition function will acquire some additional factors due to the anomaly. The exception is the partition function on a so-called canonical (spatial) circle, $Z_{1,0}(\bar{\hat{\kappa}}|\hat{\kappa})=Z_{\bar{\kappa}|\kappa}(0|-1)$.
Then the high temperature limit on the left hand side becomes a low temperature limit on the right hand side, and the dominate contribution to the later is given by the vacuum expectation values of the generators.
By keeping track of the appropriate anomalies (for details see \cite{DHH,CHI}) and taking the limit $\zeta\to\infty$, we get the dominant contribution
\begin{align}\label{canonicalZ}
Z_{0,1}(\bar{\hat{\kappa}}|\hat{\kappa})=e^{\frac{ k }{4}\frac{\bar{\hat{\kappa}}^2}{\hat{\kappa}}}Z_{0,1}(-\frac{2\pi i\bar{\hat{\kappa}}}{\hat{\kappa}}|\frac{4\pi^2}{\hat{\kappa}})=\, e^{\frac{ k }{4}\frac{\bar{\hat{\kappa}}^2}{\hat{\kappa}}}e^{-\frac{2\pi i \bar{\hat{\kappa}}}{\hat{\kappa}}P_0^{vac}-\frac{4\pi^2 }{\hat{\kappa}}L_0^{vac}}\,,
\end{align}
where $P_{0}^{vac}$ and $L_{0}^{vac}$ are the expectation values of the charges associated with $P_0$ and $L_0$ on the torus with  $({\bar a}',a')=(0,1)$ and $( \bar{\hat \kappa}',\hat{\kappa}')=( -\frac{2\pi  i\bar{\hat{\kappa}}}{\hat{\kappa}},\frac{4\pi^2}{\hat{\kappa}})\rightarrow ( {-i\alpha\zeta\over\pi}, 2\zeta)$.
 Note that when $\zeta\to\infty$, the thermal circle is infinitely long, but with a finite slope ${{ \bar{\hat \kappa}}'\over {{\hat \kappa}}'}\rightarrow {-i\alpha\over 2\pi}$.
For finite values of $\alpha$, the $\epsilon\rightarrow 0$( or equivalently $\zeta\rightarrow \infty$) limit gives a tilted cylinder. Comparing to results in the literature, taking finite $\alpha$ is in fact the slow rotating limit described in section 3.1 of \cite{DHH}.
When $\alpha=0$, the torus is going to a degenerate limit.

(\ref{canonicalZ}) is valid if the spectrum of $L_0$ is bounded from below, and the vacuum has no macroscopic degeneracy. See \cite{DHH,CHI,NBC} for discussions about the spectrum and unitary representations. An extension of the range of validity of the (\ref{canonicalZ}) is possible along the lines of \cite{Hartman-Keller}.
In general, without knowing more information of the WCFT, we do not know how to determine the value of $\alpha$, and also the vacuum charges $P_0^{vac}$ and $L_0^{vac}$.
However, for WCFTs with a holographic duals, we can deduce them from the bulk theory. We will give a precise dictionary later. Here in this section, we will keep $\alpha$, $P_0^{vac}$ and $L_0^{vac}$ undetermined.
Then  the thermal entropy is
\begin{align}\label{thermalentropy0}
S_{thermal}=(1-\hat{\kappa}\partial_{\hat{\kappa}}-\bar{\hat{\kappa}}\partial_{\bar{\hat{\kappa}}} )\log Z_{1,0}(\bar{\hat{\kappa}}|\hat{\kappa})\,.
\end{align}
One can check that the entropy is invariant under (\ref{ctcanonical}), hence (\ref{thermalentropy0}) also gives the thermal entropy of WCFT$_{(x,t)}$ defined on the ``Rindler space'', and furthermore gives the entanglement entropy of the interval (\ref{intervalXT}). Plugging (\ref{canonicalZ}) into (\ref{thermalentropy0}) we get the entanglement entropy
\begin{align}\label{SeeX}
S_{EE}(\mathcal{A})=&\frac{2  \pi  (-i \bar{\hat{\kappa}} P_0^{vac}-4 \pi L_0^{vac} )}{\hat{\kappa} }
\cr
=&-i P_0^{vac} \left(\textit{l}_T+ \frac{\bar{\beta}-\alpha}{\beta} \textit{l}_X\right)+\left(-i\frac{\alpha}{\pi} P_0^{vac}-4   L_0^{vac}\right) \log\left(\frac{\beta}{\pi\epsilon}\sinh\frac{\textit{l}_X \pi}{\beta}\right)\,.
\end{align}
(\ref{SeeX}) is our result for the entanglement entropy of an interval parameterized by $(l_X,l_T)$ in a WCFT with thermal circle $(\beta,{\bar\beta})$. $P_0^{vac}$ and $L_0^{vac}$ are the expectation values of $L_0$ and $P_0$ on tilted cylinder parameterized by $\alpha$.
We expect that the $\alpha$ in (\ref{SeeX}) is not an arbitrary parameter, and will be fixed by the theory.
This assumption will be consistent with the assumption that entropy should be invariant under the warped conformal transformations, and will also be consistent with our bulk calculations later.  When $\alpha=0$, (\ref{SeeX}) has the same expression of \cite{CHI}.

For later convenience, we also write down explicitly the combination of the coordinate transformations (\ref{warpedmapping1}) and (\ref{ctcanonical}) \begin{align}\label{warpedmapping2} \frac{\tanh\frac{\pi X}{\beta}}{\tanh\frac{\textit{l}_X \pi}{2 \beta}}=\tanh\frac{\pi \hat{x}}{\hat{\kappa}}\,, \qquad T+\left(\frac{\bar{\beta}}{\beta}-\frac{ \alpha }{\beta}\right)X=\hat{t}+\left(\frac{\hat{\bar{\kappa}}}{\hat{\kappa}}-\frac{\alpha }{\hat{\kappa}}\right) \hat{x}\,,  \end{align} which is also a warped conformal mapping satisfying (\ref{warpedmapping}). We denote the WCFT on the ($\hat{x},\hat{t}$) as  WCFT$_{(\hat{x},\hat{t})}$.

\section{The gravity side story}\label{secgravitystory}
In this section, we apply the Rindler method \cite{CHM} to the gravity side, in the context of AdS/WCFT or WAdS/WCFT.
Following  the logic of \cite{CHM}, we first find the analog of hyperbolic black holes in the bulk, and then calculate the regularized entropy. We propose that this regularized thermal entropy is just the holographic entanglement entropy for a WCFT. We leave the precise matching between the bulk and boundary calculations to next section.

The simplest model is AdS$_3$ in Einstein gravity, with the CSS boundary conditions \cite{NBC}.
We  will also consider the class of three dimensional theories of gravity obtained from consistent truncations of string theory discussed in \cite{Detournay-Guica,CGR}. Both BTZ black strings and WAdS$_3$ black strings are solutions to these theories. An interesting feature is that the thermodynamical properties and the asymptotic symmetries of all the WAdS$_3$ black strings are identical to those of BTZ after some appropriate reparametrization.
In particular, the entropy of all these black strings are given by the horizon length. In this section we will not specify what theory we are using. The discussions below apply to all the models in \cite{Detournay-Guica}, as well as BTZ black holes and black strings in Einstein gravity.

The spacetimes we use in this section are the warped black strings (\ref{blackstring}), whose asymptotic symmetry algebras are the non-canonical ones (\ref{tildealgebra}), rather than the canonical ones (\ref{wcftalg}) we used on the field side story. To do the exact matching between both sides we need to do the additional state-dependent coordinate transformation (\ref{nlct}). We will do the matching in Sec.\ref{secmatching}.

We will only explicitly write down the formulas for WAdS$_3$. However, since all the transformations and thermodynamic quantities in this section are independent from $\lambda$, by setting $\lambda=0$ these formulas also give the story for AdS$_3$ with CSS boundary conditions.

We start with  the strategy in \ref{strategy}, and then discuss the example with $T_{V}=1,\, T_{U}=0$ in section \ref{sectu0tv1}, and general temperatures in section \ref{arbitrary temperatures}.

\subsection{The strategy }\label{strategy}
We now extend the field side story into the bulk with the WAdS/WCFT correspondence.
In the WCFT calculation, the {\it Rindler transformation} (\ref{warpedmapping1}) maps entanglement entropy of WCFT$_{(X,T)}$ to a thermal entropy of WCFT$_{(x,t)}$ on an torus parametrized by (\ref{thermalcircle}),(\ref{spatialcircle}).
By (W)AdS/WCFT correspondence outlined in section 2, WCFT$_{(X,T)}$ is holographically dual to a warped black string WAdS$_{(U,V,\rho)}$.  Similarly, the bulk dual of a torus (presumably above the Hawking-Page temperature) is expected to be the (W)AdS black holes, with the spatial circle related to the horizon length and direction, and the thermal identifications related to the Hawking temperature and angular velocity.
When the spatial circle is uncompactified, we expect to find the (W)AdS black strings (W)AdS$_{(u,v,r)}$. Furthermore, the bulk coordinate transformation between (W)AdS$_{(U,V,\rho)}$ and (W)AdS$_{(u,v,r)}$ should implement the {\it Rindler transformation} (\ref{warpedmapping1})-(\ref{spectralflow}) . The thermal entropy of WCFT$_{(x,t)}$ should be the Bekenstein-Hawking entropy of WAdS$_{(u,v,r)}$.
Then the entanglement entropy for WCFT$_{X,T}$  should be calculated by the Bekenstein-Hawking entropy of WAdS$_{(u,v,r)}$.

In the gravity story, we therefore need to find out the analog of hyperbolic black holes (W)AdS$_{(u,v,r)}$ with the following conditions. Firstly, (W)AdS$_{(u,v,r)}$ satisfies a Dirichelet-Neumann boundary conditions with a spatial identifications given by (\ref{spatialcircleuvr}), and the temperature and angular velocity should be related to the thermal identification (\ref{tcuv}). Secondly,  (W)AdS$_{(u,v,r)}$ is a warped black string in the form of (\ref{blackstring}), which has two commuting Killing vectors. Thirdly, the coordinate transformation between (W)AdS$_{(U,V,\rho)}$ and (W)AdS$_{(u,v,r)}$ at the boundary should be the {\it Rindler transformation} (\ref{warpedmapping1})-(\ref{spectralflow}).

Our strategy is to use the quotient method to find WAdS$_{(u,v,r)}$, which is the bulk dual of WCFT$_{(x,t)}$. Note that WAdS$_{(U,V,\rho)}$ has local isometry $SL(2,R)_R\times U(1)_L$, let us denote the generators of $SL(2,R)$ by $J_{0,\pm}$, and the generator of $U(1)$ by $J_L$. We expect the WAdS$_{(u,v,r)}$ to have two explicit commuting Killing vectors. One Killing vector must be proportional to $J_L$, while the other must be a linear combination of the four generators. We could further require that the new radial direction is orthogonal to the two Killing vectors. Solving all these conditions we will get the coordinate transformation, as well as the new metric, up to a reparametrization of $r$, and  some integration constants. This method was used to build and classify locally AdS$_3$ solutions \cite{BHTZ}, and was generalized in
\cite{warpedblackholes} to find WAdS$_3$ black hole solutions.

Here we briefly comment on AdS$_3$. For AdS$_3$, the local isometry is $SL(2,{\mathbb R})_R\times SL(2,{\mathbb R})_L$. With CSS boundary conditions, however, only the $SL(2,{\mathbb R})_R\times U(1)_L$ part belongs to the asymptotic symmetries. To get a warped conformal transformation at the boundary,  the bulk coordinate transformation should be built from the $SL(2,{\mathbb R})_R\times U(1)_L$ quotient. Alternatively, if the Killing vectors of the new metric are from the entire $SL(2,{\mathbb R})_R\times SL(2,{\mathbb R})_L$ generators, we will implement a conformal transformation on the boundary. Corresponding, we will get a result compactable with the AdS/CFT correspondence.  The {\it Rindler transformation} in the bulk should be allowed by the boundary conditions. The different ways of doing quotient for AdS$_3$  (as elaborated in Appendix \ref{appendix2}) show that the boundary conditions play an essential role in holographic entanglement entropy.

\subsection{The story with $T_{V}=1,\, T_{U}=0$}\label{sectu0tv1}
In this subsection, we will elaborate the ideas for a simple example of  WAdS black string with $T_{V}=1,\, T_{U}=0$. We first show how to find the coordinate transformation by the quotient method in section \ref{4.1.1}, then calculate the thermal entropy of the resulting black hole in section \ref{4.1.2}, and finally discuss the geometric quantity that captures the holographic entanglement entropy in section \ref{secgravitystorygeometricquantity}.

\subsubsection{The quotient: from WAdS$_{(U,V,\rho)}$ to WAdS$_{(u,v,r)}$}\label{4.1.1}
We start with a warped black string WAdS$_{(U,V,\rho)}$ with $T_ V=1,T_ U=0$
\begin{align}\label{wbsT1T0}
ds^2= \ell^2\Big(\frac{\left(\lambda ^2+1\right) d \rho^2}{4  \rho^2}-\lambda ^2  \rho^2 d U^2+2  \rho d U d V+d V^2\Big)\,,
\end{align}
with Killing vectors
\begin{align}\label{KVs}
J_L=&\partial_ V\,,
\cr
J_+=&\frac{4  \rho^2  U^2+1}{4  \rho^2}\partial_ U-\frac{1}{2  \rho}\partial_ V-2  \rho  U\partial_ \rho\,,
\cr
J_0=& U\partial_ U- \rho\partial_ \rho\,,
\cr
J_-=&\partial_ U\,,
\end{align}  where the normalization are chosen to satisfy the standard $SL(2,R)$ algebra
$[J_-,J_+]=2J_0\,,\, [J_0,J_\pm]=\pm J_{\pm}$.
Define  \begin{align}\label{Jgeneral}
J=a_L J_L+a_0 J_0+a_+ J_++a_- J_-\,,
\end{align}
where
 $a_0,a_+,a_-,a_L$ are arbitrary constants.
We define some new coordinates such that there are two explicit Killing vectors \be  \p_ u = J\,, \quad \p_ v =J_L\,, \label{uvdef}\ee
From (\ref{uvdef}) we get some components of the new metric
\be g_{uu}=J \cdot J\,, \quad g_{uv}=J\cdot J_L\,, \quad g_{vv}=J_L\cdot J_L\label {guv}\,,\ee
where the inner products are calculated with the old metric for WAdS$_{(U,V,\rho)}$.
The new metric should only depend on the third coordinate $r$.  Up to reparametrization, we can always choose the new radial coordinates by
\be r\equiv{J\cdot J_L\over \ell^2}\,. \label{radial}\ee
It is easy to verify that $g_{uu}$ and $g_{vv}$ only depend on $r$.
We further require that the new metric has no cross terms between $r$ and $u,v$, namely
\bea
g_{ru}=n_r\cdot J=0, \quad g_{rv}=n_r\cdot J_L=0,\quad n_r\equiv \p_r\,. \label{orth}
\eea
Solving all these conditions will give the coordinate transformation, as well as the new metric.

 The most general quotient with arbitrary parameters $a_0,a_+,a_-,a_L$ is given in Appendix \ref{appendix1}. The boundary coordinates $(u,v)$ covers a strip of $(U,V)$, with
 $a_0/(2a_+)$ controlling the center position of  strip, while $-\sqrt{a_0^2-4 a_- a_+}/a_+$ controlling the width of the strip. It turns out that the regularized Bekenstein-Hawking entropy of the WAdS$_{(u,v,r)}$ only depends on the width of the strip. Hence, for simplicity but without losing generality, we choose the parameters in the main text as follows
\begin{align}\label{ala0a+a-}
a_L= a_0=0\,, \quad a_+=-{2 \over l_U},\quad a_-={l_U\over2 }\,.
 \end{align}
With the above choice of the parameters, we find the following coordinate transformation
\begin{align}\label{ctUu}
 u=&\frac{1}{4  } \log \left(\frac{(l_U +2 U)^2\rho^2-1}{(l_U -2 U)^2\rho^2- 1}\right)\,,
\cr
 v =&\frac{1}{4} \log \left(\frac{(1+2 \rho U)^2-l_U ^2 \rho^2}{(1-2\rho U)^2- l_U ^2 \rho^2}\right)+V\,,
\cr
 r =&\frac{ 1+\rho^2 \left(l_U ^2-4U^2 \right)}{2 l_U\rho}\,,
\end{align}
Under which we get a new warped black string,  denoted by WAdS$_{(u,v,r)}$,
\begin{align}\label{bhmetric}
ds^2&=\ell^2\left(\left(T_{ u }^2\left(1+\lambda ^2T_{ v }^2 \right) -\lambda ^2 r^2\right)\,d u ^2+2  r   \,d u  d v + T_{ v }^2d v ^2+ \frac{\left(1+\lambda ^2T_{ v }^2\right) }{4 ( r ^2 -T_{ u }^2T_{ v }^2)}d r  ^2\right)\,,\\
T_{ u }&=T_{ v }=1\,,
\end{align}
with an infinite event horizon at \be r_h=T_u T_v\,.\ee
It is easy to verify that WAdS$_{(u,v,r)}$ (\ref{bhmetric}) and WAdS$_{(U,V,\rho)}$ (\ref{wbsT1T0}) satisfy the Dirichlet-Neumann boundary conditions (\ref{css1}), (\ref{css}) with the same $T_v$ but with different spatial identification.

\subsubsection{A bulk calculation of the entanglement entropy}\label{4.1.2}
Similar to the story of AdS$_3$/CFT$_2$, the WAdS$_3$/WCFT dictionary will translates the bulk calculation to a boundary calculation.
As discussed in section 2,   asymptotic symmetry analysis directly relates gravity on WAdS$_{(U,V,\rho)}$ to WCFT$_{(U,V)}$ with the tilded algebra (\ref{tildealgebra}).

Let us look at the coordinate transformations on the boundary, which are given by
\begin{align}\label{ctUub}
 u =&{1\over 2 }\log \left(\frac{l_U +2 U}{l_U -2 U}\right)+\mathcal{O}(\frac{1}{\rho^2})\,,\\
 \label{ctUub1}
 v =&V+\mathcal{O}(\frac{1}{\rho})\,.
\end{align}
The boundary coordinate transformations (\ref{ctUub}) and (\ref{ctUub1}) indicate that the boundary of WAdS$_{(u,v,r)}$ covers a strip with $-{ l_U  \over2}<U<{ l_U  \over2} $, on the boundary of WAdS$_{(U,V,\rho)}$.

Similar to the discussion on the field theory side, now let us  consider an interval
\begin{align}\label{intervalUV'1}
\mathcal{A}:~~\{(U,\,V)|\, U={(l_U  -2 \epsilon)}\left(-{1\over2}+\tau\right) ,\quad V={l_V}\left(-{1\over2}+\tau\right),\quad \tau\in[0,1]\}\,,
\end{align}
with $\epsilon$ being a small cutoff. We find that on the boundary of WAdS$_{(u,v,r)}$, the interval (\ref{intervalUV'1}) in terms of the new coordinates ($ u , v $) is given by
\begin{align}\label{intervaltutv}
\{(u,\,v)|\, u={\Delta u}\left(-{1\over2}+\tau\right) ,\quad v={\Delta v}\left(-{1\over2}+\tau\right),\quad \tau\in[0,1]\}\,,
 \end{align}
with
\begin{align}\label{dudv}
\Delta  u =\log\left(\frac{ l_U}{ \epsilon}\right)\,,\qquad \Delta  v =l_V\,.
\end{align}
We see the interval (\ref{intervaltutv}) is infinitely extended in the $ u $ direction as $\epsilon\rightarrow 0$. As in section \ref{2.2.3}, we identify the end points of the interval. Thus, beside the thermal circle,
we also have a spatial circle in WAdS$_{(u,v,r)}$
\begin{align}\label{spatialcircletutv}
spatial \,circle: ( u , v )\sim( u +\Delta  u , v +\Delta  v )\,,
\end{align}

According to our discussions in section \ref{secblackstring}, given the temperatures $T_u=T_v=1$,  the total thermal entropy of WAdS$_{(u,v,r)}$(\ref{bhmetric}) on the interval (\ref{intervaltutv})  is given by (\ref{suv}). In terms of variables of WAdS$_{(U,V\rho)}$ and using (\ref{dudv}), we get
\begin{align}\label{BHentropy01}
S_{HEE}
=&{\ell\over4G}l_V+{\ell\over4G}\log\frac{l_U}{\epsilon}\,.
\end{align}
We propose that  (\ref{BHentropy01}) is the bulk calculation for the entanglement entropy in the contexts of WAdS/WCFT and AdS/WCFT.
We will show explicitly how  (\ref{BHentropy01}) reproduce the WCFT result (\ref{SeeX}) in next section.

\subsubsection{The geometric quantity in WAdS$_{(U,V,\rho)}$}\label{secgravitystorygeometricquantity}
\begin{figure}

\centering
\includegraphics[width=0.9 \textwidth]{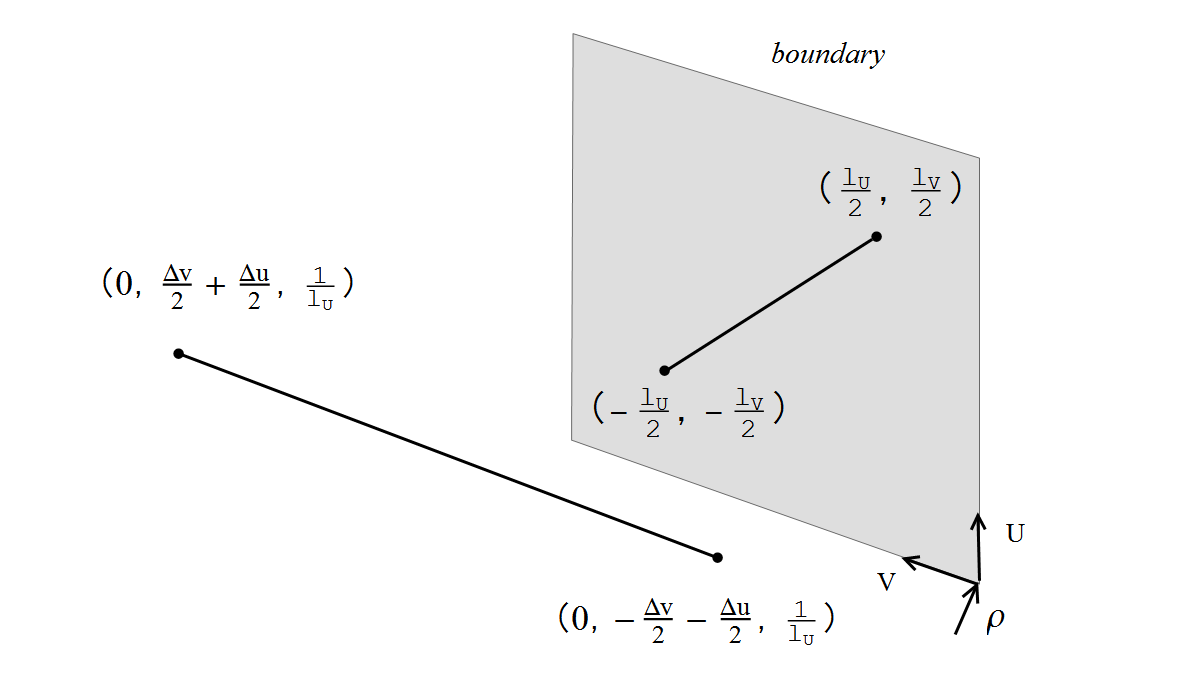}

\caption{\label{Fig3} The solid line segment in the bulk is the geometric quantity $\gamma_{\mathcal{A}}$ (\ref{image1}) whose length is proportional to the entanglement entropy of the interval on the boundary (\ref{intervalUV'1}). }
\end{figure}

We can also calculate the inverse coordinate transformations from WAdS$_{(u,v,r)}$ to WAdS$_{(U,V,\rho)}$, and find out what is the image of the horizon of WAdS$_{(u,v,r)}$. The inverse coordinate transformations are
{\footnote{There are two branches of the inverse coordinate transformations and both satisfy (\ref{ctUu}). The other branch is given by
\begin{align}
U= \frac{l_U \sqrt{ r ^2-1} \sinh \left(2   u \right)}{2 \left(\sqrt{ r ^2-1} \cosh \left(2  u \right)- r \right)}\,,\qquad \rho=& \frac{ \left( r -\sqrt{ r ^2-1} \cosh \left(2  u \right)\right)}{ l_U}\,.
\end{align}
We drop this branch since it will give negative $\rho$ when $ u $ is big enough.}},
\begin{align}\label{inverseRindler}
U&= \frac{l_U \sqrt{ r ^2-1} \sinh \left(2   u \right)}{2 \left(\sqrt{ r ^2-1} \cosh \left(2  u \right)+ r \right)}\,,
\\
\rho&= \frac{ \left( r +\sqrt{ r ^2-1} \cosh \left(2  u \right)\right)}{ l_U}\,,
\\
\label{inverseV}
V&=\frac{1}{4}\log \left(\frac{2 \sqrt{ r ^2-1}  e^{2  u }+\left( r +1\right) e^{4  u }+ r -1}{2  \sqrt{ r ^2-1} e^{2  u }+\left( r -1\right) e^{4  u  }+ r +1}\right)+ v\,.
\end{align}
We parametrize the horizon of WAdS$_{(u,v,r)}$ in the following way
\begin{align}\label{wbhhorizon}
 \{(u,\,v,\,r)|\,u =-\frac{\Delta  u }{2}+\tau \Delta u \,,\quad  v =-\frac{\Delta  v }{2}+\tau \Delta v \,,\quad  r =1,\quad \tau\in[0,1]\}\,.
\end{align}
The inverse coordinate transformations (\ref{inverseRindler})-(\ref{inverseV}) indicate that the image interval of the horizon (\ref{wbhhorizon}), denoted by $\gamma_{\mathcal A}$, is an interval with both the $\rho$ and $U$ coordinates fixed (See Fig.\ref{Fig3}), depicted by
\begin{align}\label{image1}
\gamma_{\mathcal A}:~~\{(U_h,\,V_h,\,\rho_h)|\, U_h=0 ,\quad \rho_h=\frac{1}{l_U},\quad V_h=(\Delta u+\Delta v)\left(-\frac{1}{2}+\tau\right),\quad \tau\in[0,1]\}\,.
\end{align}
Using (\ref{dudv}) and $g_{VV}=\ell^2$, we find the proposed holographic entanglement entropy (\ref{BHentropy01}) is given by the length of $\gamma_{\mathcal A} $
\begin{align}
S_{HEE}={Length(\gamma_{\mathcal A}) \over 4G}\,,
\end{align}

Note that $\gamma_{\mathcal{A}}$ is a geodesic, but not anchored on the end points of the interval (\ref{intervalUV'1}) on the asymptotic boundary. This is the main difference between this bulk calculation in WAdS and the RT proposal for AdS  \cite{RT1,RT2}. It was noticed that in \cite{SWX} that boundary conditions play a role in the Lewkowycz-Maldacena \cite{LM} derivation of holographic entanglement entropy. It will be interesting to see how to modify the Lewkowycz-Maldacena \cite{LM} prescription in this case. This may give us a better understanding of the curve $\gamma_{A}$. We leave this for future investigations.

\subsection{The story with arbitrary temperatures}\label{arbitrary temperatures}
In this subsection we extend  our discussions to WAdS$_{(U,V,\rho)}$ with arbitrary temperatures
\begin{align}\label{wbsgeneralT}
ds^2=\ell^2 \left(\left(T_U^2 \left(\lambda ^2 T_V^2+1\right)-\lambda ^2 \rho^2\right)dU^2+2 \rho dU dV+T_V^2 dV^2+\frac{ \left(\lambda ^2 T_V^2+1\right)d\rho^2}{4 \left(\rho^2-T_U^2 T_V^2\right)}\right)\,.
\end{align}
Similar to the case of $T_V=1,T_U=0$, we can directly use the quotient method following the steps in section \ref{sectu0tv1}.
Alternatively we can first perform a coordinate transformation to reduce the metric (\ref{wbsgeneralT}) of WAdS$_{(U,V,\rho)}$ to the metric (\ref{wbsT1T0}) with $T_U=0,T_V=1$,  and then use the result in section \ref{sectu0tv1}. This amounts to performing the following coordinate transformation
\begin{align}\label{ctUu'}
&4 T_U U =\log \left(\frac{ \left(\sqrt{r^2-T_{V}^2} \left(\cosh (2 u )+\sinh (2 u ) \tanh \left(\frac{l_U  T_U}{2}\right)\right)+r\right){}^2-T_{V}^2 \tanh ^2\left(\frac{l_U  T_U}{2}\right)}{ \left(\sqrt{r^2-T_{V}^2} \left(\cosh (2 u )-\sinh (2 u ) \tanh \left(\frac{l_U  T_U}{2}\right)\right)+r\right){}^2- T_{V}^2\tanh ^2\left(\frac{l_U  T_U}{2}\right)}\right)\,,
\cr
&2 T_V V = 2 T_V v+\coth ^{-1}\left(\frac{\text{csch}(2 u )}{T_{V}} \left(\sqrt{r^2-T_{V}^2}+r \cosh (2 u )\right)\right) -
 \cr
 & \tanh^{-1}\left(\frac{4 T_{V}\sqrt{r^2-T_{V}^2} \sinh (2 u ) \tanh ^2\left(\frac{l_U   T_U}{2}\right)}{\text{sech}^2\left(\frac{l_U   T_U}{2}\right) \left(\left(r^2-T_{V}^2\right) \cosh (4 u )-r^2+3T_{V}^2\right)+4 r \sqrt{r^2-T_{V}^2} \cosh (2 u )+4 \left(r^2-T_{V}^2\right)}\right)\,,
\cr
&~~~~\rho =T_U  \text{csch}\left(l_U  T_U\right) \left(r \cosh \left(l_U  T_U\right)+\sqrt{r^2-T_{V}^2} \cosh (2 u )\right)\,.
\end{align}
Under (\ref{ctUu'}) we get the metric of WAdS$_{(u,v,r)}$
\begin{align}\label{bhmetric'}
\frac{ds^2}{\ell^2}&=\left(T_{ u }^2\left(1+\lambda ^2T_{ v }^2 \right) -\lambda ^2 r^2\right)\,d u ^2+2  r   \,d u  d v + T_{ v }^2d v ^2+ \frac{\left(1+\lambda ^2T_{ v }^2\right) }{4 ( r ^2 -T_{ u }^2T_{ v }^2)}d r  ^2\,,
\\
T_{ u }&=1\,,\qquad T_{ v }=T_{V}\,.
\end{align}
The coordinate transformations on the boundary are given by
\begin{align}\label{ctUub'}
\frac{\tanh\left( T_U U \right)}{\tanh \left(\frac{l_U  T_U}{2}\right)}=&\tanh (u ) +\mathcal{O}(\frac{1}{r^2})\,,\\
\label{ctUub'1}
V =&v+\mathcal{O}(\frac{1}{r})\,,
\end{align}
which similarly indicate that the boundary of WAdS$_{(u,v,r)}$ covers a strip on the boundary WAdS$_{(U,V,\rho)}$ of with width ${\textit{l}_U}$.

Again we consider an interval (\ref{intervalUV'1}),
and find that on the boundary of WAdS$_{(u,v,r)}$, the interval (\ref{intervalUV'1}) in terms of the new coordinates ($ u , v $) is given by (\ref{intervaltutv}), but with
\begin{align}\label{dudvgeneralT}
\Delta u=\log \left(\frac{\sinh (l_U T_U)}{\epsilon T_U }\right)\,,\qquad
\Delta v=l_V\,.
\end{align}
Again this interval (\ref{intervaltutv}) is infinitely extended along the $ u $ direction, and hence we identify the end points, which lead to a spatial circle
\begin{align}\label{spatialcircletutv}
spatial \,circle: ( u , v )\sim( u +\Delta u , v +\Delta v )\,,
\end{align}
According to our discussions in Sec. \ref{secblackstring}, the thermal entropy of WAdS$_{(u,v,r)}$ (\ref{bhmetric'}) with the above spatial circle is given by
\begin{align}\label{BHentropy}
S_{HEE}=&\frac{\ell}{4G}\left( T_V\textit{l}_V+\log \left(\frac{\sinh (l_U T_U)}{\epsilon T_U }\right)\right)\,.
\end{align}

We can also find the image interval $\gamma_{\mathcal{A}}$ of the horizon of  WAdS$_{(u,v,r)}$
\begin{align}\label{image2}
\{(U_h,\,V_h,\,\rho_h)|\, U_h=0 ,\, \rho_h=T_U T_V \coth \left(l_U T_U\right),\, V_h=\left(\frac{T_{V} \Delta v+\Delta u }{ T_V}\right)\left(-\frac{1}{2}+\tau\right),\, \tau\in[0,1]\}\,.
\end{align}
Using (\ref{dudvgeneralT}) and $g_{VV}=\ell^2 T_V^2$, we find the length of $\gamma_{\mathcal{A}}$ is given by
\begin{align}
Length(\gamma_{\mathcal{A}})=\ell\left( T_V\textit{l}_V+\log \left(\frac{\sinh (l_U T_U)}{\epsilon T_U }\right)\right)\,,
\end{align}
and also
\begin{align}
S_{HEE}={Length(\gamma_{\mathcal A}) \over 4G}\,.
\end{align}

\section{Matching between WAdS$_3$ and the WCFT}\label{secmatching}
In this section, we relate the entanglement entropy in the WCFT (\ref{SeeX}) to the bulk calculation (\ref{BHentropy}). Section \ref{5.1} relates the WCFT$_{(U,V)}$, WCFT$_{(u,v)}$ and WCFT$_{(\hu,\hv)}$ appearing from the bulk analysis to a WCFT$_{(X,T)}$ , WCFT$_{(x,t)}$ and WCFT$_{(\hat x,\hat t)}$ with canonical algebra. Section \ref{5.2} shows that  the bulk coordinate transformation is indeed an extension of the warped conformal transformation. Section \ref{5.3} shows how to determine the vacuum charges using holography. Section  \ref{5.4} finally shows that the bulk result and the WCFT result agree with each other.

\subsection{WAdS$_{(U,V,\rho)}$, WCFT$_{(U,V)}$  and WCFT$_{(X,T)}$}\label{5.1}
As discussed in Sec. \ref{2.2}, the asymptotic symmetry of AdS$_3$ and WAdS$_3$ under the Dirichlet-Neumann boundary conditions is given by the algebra (\ref{tildealgebra}) \cite{DHH,Compere-Detournay,Detournay-Guica,CGR}.
Given a WAdS$_{(U,V,\rho)}$ with arbitrary temperatures $T_U$ and $T_V$ (\ref{wbsgeneralT}), the boundary theory WCFT$_{(U,V)}$ is naturally defined with coordinates $U$ and $V$.
The algebra (\ref{tildealgebra}) is related to the canonical WCFT algebra (\ref{wcftalg}) by a redefinition of the generators. In particular, for states with vanishing  $P_{n\neq0}$, the mapping can be achieved by a (state-dependent) coordinate transformation (\ref{nlct}),
\begin{align}\label{ctXU}
U=X\,, \qquad V=\frac{1}{T_V}\sqrt{\frac{-Gk}{\ell}} T+X\,,
\end{align}
We will denote the WCFT in the $(X,T)$ coordinates by WCFT$_{(X,T)}$.
The thermal circle of the black string (\ref{wbsgeneralT}) is
\be
thermal\, circle: \quad (U,V) \sim \left(U+{\pi i\over T_U},~ V-{\pi i\over T_V}\right)\,, \label{tcUV}
\ee
Under the coordinate transformation (\ref{ctXU}), the above thermal circle (\ref{tcUV}) induces a thermal
circle for WCFT$_{(X,T)}$
\bea \label{tcXT}
thermal\, circle:\quad (X,T)&\sim& (X+i\beta,T-i\bar{\beta})\,,\\
\beta&=&\frac{\pi}{T_U}\,,\qquad \bar{\beta}=\pi\sqrt{\frac{-\ell}{Gk}}  \left(\frac{T_V}{T_U}+1\right)\,.
 \label{beta2TU}
\eea
Similarly, a spatial interval on the $(U,V)$ plane parameterized by $l_U,\,l_V$
will be mapped to a spatial interval with
\begin{align}\label{dTdX} spatial\, interval:\quad \{(X,\,T)|&\, X={l_X }\left(-{1\over2}+\tau\right) ,\quad T={l_T}\left(-{1\over2}+\tau\right),\quad \tau\in[0,1]\}\,,
\\\label{dX2dU}
\textit{l}_T=&T_V\sqrt{\frac{-\ell}{Gk}}(\textit{l}_V- \textit{l}_U),\qquad \textit{l}_X=\textit{l}_U\,.
\end{align}
According to the discussions in section \ref{secfeildtheorystrory}, the entanglement entropy for a spatial interval (\ref{dTdX}) on the WCFT$_{(X,T)}$  with a thermal circle (\ref{tcXT}) is given by (\ref{SeeX}).

\subsection{WAdS$_{(u,v,r)}$, WCFT$_{(u,v)}$ and WCFT$_{(x,t)}$
}\label{5.2}
In this subsection we map WCFT$_{(u,v)}$, obtained from WAdS$_{(u,v,r)}$, to WCFT$_{(x,t)}$, which will relate the bulk coordinate transformation (\ref{ctUu'}) and the warped conformal transformation (\ref{warpedmapping1})-(\ref{spectralflow}).

Similar to the previous subsection, we get WCFT$_{(u,v)}$ from asymptotic analysis of WAdS$_{(u,v,r)}$.
The transformation \begin{align} \label{ctux}
u=x\,,\quad v=\sqrt{\frac{-Gk}{\ell}}\frac{t}{T_{v}}+x\,,
\end{align}
leads to a torus
\begin{align}
spatial:&~(x,t)\sim \left(x+\Delta u,t+T_v\sqrt{\frac{-\ell}{Gk}}(\Delta v-\Delta u)\right)\,,
\cr
thermal:&~(x,t)\sim \left(x+i\frac{\pi}{T_u},t-i \pi \sqrt{\frac{-\ell}{Gk}}\left(\frac{T_v}{T_u}+1\right)\right)\,.
\label{torusxt}
\end{align}
Note that the  bulk coordinate transformation (\ref{ctUu'}) induces a warped conformal mapping from WCFT$_{(U,V)}$ to WCFT$_{(u,v)}$, given by (\ref{ctUub'}).
 (\ref{ctUub'}) can be rewritten in terms of variables in the $(X,T)$ and  $(x,t)$ coordinates using (\ref{ctXU}) and (\ref{ctux})
\begin{align}\label{ctXx}
\frac{\tanh\frac{\pi X}{\beta}}{\tanh\frac{\textit{l}_X \pi}{2 \beta}}=\tanh{T_ux}\,,
\qquad
T+T_V \sqrt{\frac{-\ell}{Gk}} X=t+T_v\sqrt{\frac{-\ell}{Gk}} x\,,
\end{align}
where we have used $ T_v=T_V.$
Comparing  to the warped conformal transformation (\ref{warpedmapping1}),
 it is easy to read the three parameters
\begin{align}\label{kappabkappa}
\kappa&=\frac{\pi}{T_u}\,,\,{\bar \kappa}=\pi \sqrt{\frac{-\ell}{Gk}}\left(\frac{T_v}{T_u}+1\right)\,,\\
\alpha&=\sqrt{\frac{-\ell}{Gk}}\pi.\label{matching}
\end{align}

Note that by choosing arbitrary $a_+a_-$ and  $a_L$ in (\ref{Jgeneral}), we can also get arbitrary $T_u$ and $T_v$. This is consistent with the fact $\kappa$ and $\bar{\kappa}$ are arbitrary and will not affect the entanglement entropy.
On the other hand, $\alpha$ shows up explicitly in the entanglement entropy, and (\ref{matching}) is  the matching condition between the bulk and the boundary. As a consistent check, we can rewrite the torus (\ref{torusxt}) in terms of the $x,t$ variables by using(\ref{dudvgeneralT}), (\ref{beta2TU}) and (\ref{dX2dU}), it is straight forward to see that indeed (\ref{torusxt}) agrees with (\ref{spatialcircle}) and (\ref{thermalcircle}). Thus we  proved that the bulk coordinate transformations (\ref{ctUu'}) is indeed a bulk extension of the warped conformal mapping (\ref{warpedmapping1}) with the choice (\ref{kappabkappa}) and (\ref{matching}) on the field theory side.

\subsection{WAdS$_{(\hu,\hv,\hat r)}$, WCFT$_{(\hu,\hv)}$ and WCFT$_{(\hat x,\hat t)}$ }\label{5.3}

In this subsection, we map WCFT$_{(\hu,\hv)}$ to WCFT$_{(\hat x,\hat t)}$. This will fix $P_0^{vac}$ and $L_0^{vac}$ in the formula of entanglement entropy (\ref{SeeX}). Meanwhile we will further check the consistency of all transformations.

As discussed in section \ref{2.2.3},  we apply the following transformation to the WAdS$_{(u,v,r)}$ metric (\ref{bhmetric'}),
\begin{align}
&u=\frac{\Delta u}{2\pi}\hat{ u}\,,\quad v=\frac{\Delta v}{2\pi}\hat{v}\,,\quad r=\frac{4\pi^2}{\Delta v\Delta u}\hat{r}\,,
\\
\label{tuhtvh}
&\frac{\Delta u}{2\pi}=T_{\hat{u}}\,,\qquad \frac{\Delta v}{2\pi }T_{v}=T_{\hat{v}}\,,\qquad \lambda \frac{2\pi}{\Delta v}=\hat{\lambda}\,.
\end{align}
and get WAdS$_{(\hu,\hv,\hat r)}$
\begin{align}\label{wadshatu}
ds^2=&\left(T_{ \hat{u}}^2\left(1+\hat{\lambda} ^2T_{ \hat{v}}^2 \right) -\hat{\lambda} ^2 \hat{r}^2\right)\,d \hat{u}^2+2  \hat{r}  \,d \hat{u} d \hat{v}+ T_{ \hat{v}}^2d \hat{v}^2+ \frac{\left(1+\hat{\lambda} ^2T_{ \hat{v}}^2\right) }{4 ( \hat{r}^2 -T_{ \hat{u}}^2T_{ \hat{v}}^2)}d \hat{r} ^2\,,
\end{align}
with a spatial circle
\begin{align}
spatial:~(\hat{u},\hat{v})\sim (\hat{u}+2\pi,\hat{v}+2\pi)\,.
\end{align}
With the Dirichlet-Newmann type of boundary conditions, the holographic dual is WCFT$_{(\hu,\hv)}$ we discussed in section (\ref{adswcft}).
Following section (\ref{2.2.2}), a state-dependent transformation
\begin{align}
 \label{cthatx}
\hat{u}=\hat{x}\,,\quad \hat{v}=\frac{\hat{t}}{T_{\hat{v}}}\sqrt{\frac{-Gk}{\ell}}+\hat{x}\,,
\end{align}
maps  WCFT$_{(\hat{u},\hat{v})}$ to WCFT$_{(\hat{x},\hat{t})}$.

Now we double check that the connection between the bulk and boundary transformations. From (\ref{cthatx}), we get WCFT$_{(\hat{x},\hat{t})}$ on the canonical torus
\begin{align}
spatial:&~(\hat{x},\hat{t})\sim (\hat{x}+2\pi,\hat{t})\,,
\cr
thermal:&~(\hat{x},\hat{t})\sim \left(\hat{x}+i\frac{2\pi^2}{\zeta},\hat{t}-i\left(\pi\sqrt{\frac{-\ell}{Gk}}+\pi{l_T+\left(\frac{\bar \beta}{\beta}-\sqrt{\frac{-\ell}{Gk}}\frac{\pi}{ \beta}\right)l_X\over\zeta}\right)\right)\,,
\label{torushxht}
\end{align}
where (\ref{dudvgeneralT}) and (\ref{dX2dU}) are used.
Comparing (\ref{torushxht}) with (\ref{hkappa1}), we again get a matching with the condition (\ref{matching}).

The boundary coordinate transformation from $(U,V)$ to $(\hat{u},\hat{v})$ is given by
\begin{align}\label{ctUhub}
\frac{\tanh\left( T_U U \right)}{\tanh \left(\frac{l_U  T_U}{2}\right)}=&\tanh (\frac{\Delta u}{2\pi}\hat{u} )\,,\qquad  V =\frac{\Delta v}{2\pi}\hat{v}\,.
\end{align}
Using (\ref{ctXU}), (\ref{cthatx}), (\ref{dudvgeneralT}), (\ref{beta2TU}) and (\ref{dX2dU}), we get a transformation between ${X,T}$ and ${(\hat{x},\hat{t})}$
\begin{align}\label{ctXhx}
\frac{\tanh\frac{\pi X}{\beta}}{\tanh\frac{\textit{l}_X \pi}{2 \beta}}=\tanh\frac{\pi \hat{x}}{\hat{\kappa}}\,,
\qquad
T+\left(\frac{\bar{\beta}}{\beta}-\sqrt{\frac{-\ell}{Gk}}\frac{\pi }{\beta}\right)X=\hat{t}+\left(\frac{\hat{\bar{\kappa}}}{\hat{\kappa}}-\sqrt{\frac{-\ell}{Gk}}\frac{\pi }{\hat{\kappa}}\right) \hat{x}\,.
\end{align}
which is just the warped conformal mapping (\ref{warpedmapping2}) with $\alpha=\sqrt{\frac{-\ell}{Gk}}\pi \label{alphas}${}. Thus we  again proved that the bulk coordinate transformations (\ref{ctUu'}) is indeed a bulk extension of the warped conformal mapping (\ref{warpedmapping2}) with $\alpha=\sqrt{\frac{-\ell}{Gk}}\pi \label{alphas} ${} on the field theory side.

\subsection{Matching the entanglement entropy}\label{5.4}

Now we recite results from both sides, and show the matching.
On the gravity side, the proposed holographic entanglement entropy is given by (\ref{BHentropy})
\begin{align}\label{SeeUU}
S_{HEE}
=&\frac{\ell}{4 G}\, T_V \textit{l}_V+\frac{\ell}{4 G}\log\left(\frac{\sinh\left(\textit{l}_U T_U\right)}{\epsilon T_U}\right)\,,
\end{align}
while the entanglement entropy from WCFT analysis is given by   (\ref{SeeX}),
\begin{align}
S_{EE}=&-i P_0^{vac} \left(\textit{l}_T+ \frac{\bar{\beta}-\alpha}{\beta} \textit{l}_X\right)+\left(-i\frac{\alpha}{\pi} P_0^{vac}-4   L_0^{vac}\right) \log\left(\frac{\beta}{\pi\epsilon}\sinh\frac{\textit{l}_X \pi}{\beta}\right).
\end{align}
Using the vacuum values of the charges (\ref{P0L0vac}), which we rewrite here \be P_0^{vac}=\frac{i}{4}\sqrt{-\frac{\ell k}{G}},\quad L_0^{vac}=0\,.\ee{} and the matching conditions (\ref{beta2TU}), (\ref{dX2dU}),  (\ref{matching}),  it is straightforward to verify that indeed the bulk result agrees with the boundary result (\ref{SeeX})
\be S_{HEE}=S_{EE}.\ee

\begin{figure}
\centering
\includegraphics[width=1 \textwidth]{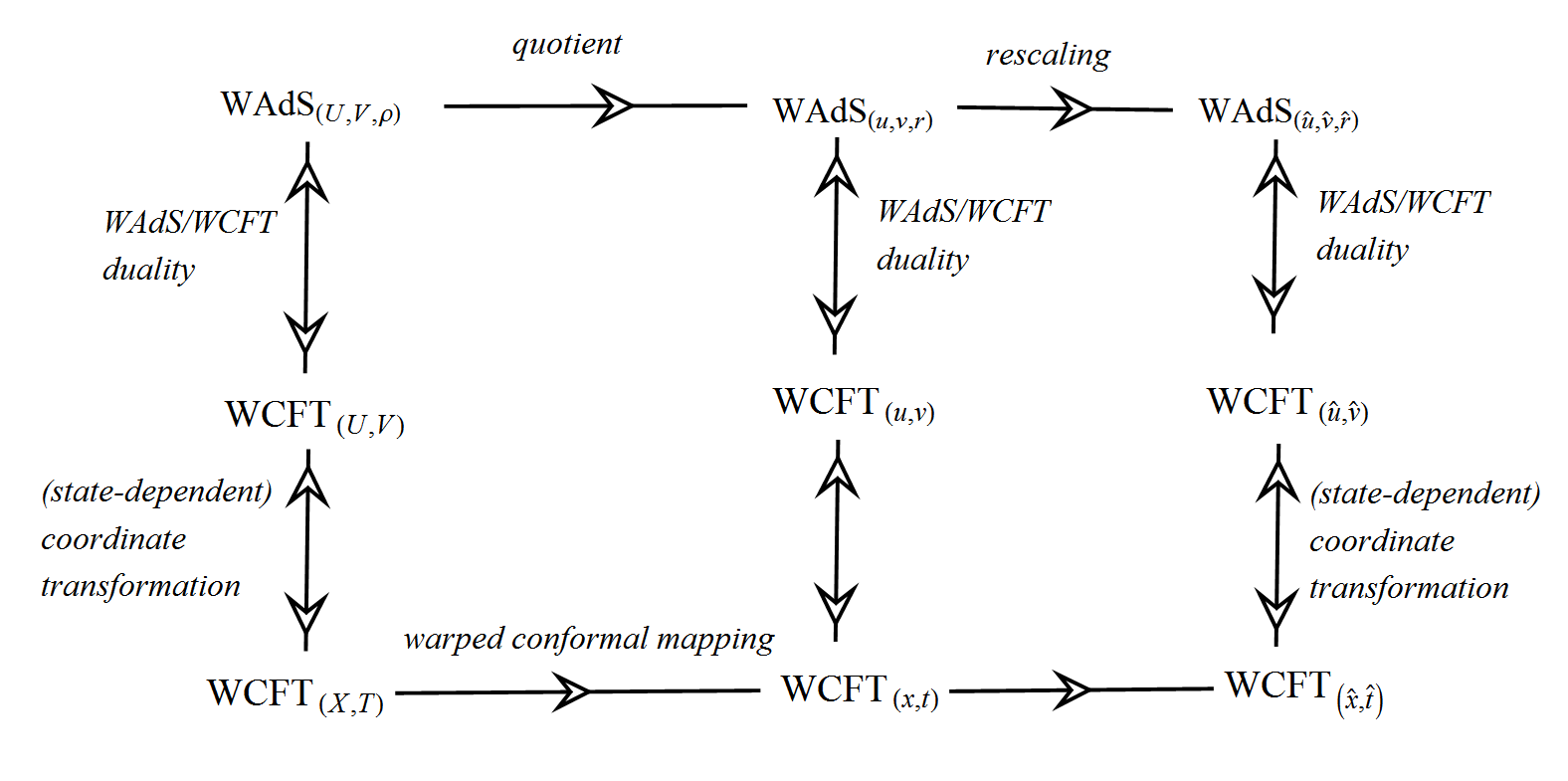}
\caption{\label{Fig1} Diagram that make a conclusion about all the spacetimes and field theories we have discussed, and their relationships. }
\end{figure}

To recapitulate,  our main steps are depicted in Fig.\ref{Fig1}, where the top line shows the calculation on the gravity side and the bottom line shows the calculation on the field theory side.
The vertical arrows show how to relate the bulk calculation to the field theory calculation.

It seems that the $(x,t)$ coordinate system in the bulk is more convenient for matching with the WCFT.  However, as there is so far no asymptotic symmetry analysis for WAdS$_{(x,t)}$, the holographic dictionary between the WAdS$_{(x,t,r)}$ and WCFT$_{(x,t)}$ is not well established directly. Nevertheless, the bulk computations for the entanglement entropy can be repeated in the $(x,t)$ coordinates directly, see Appendix C.

\section{R\'{e}nyi entropy}\label{renyi}

\subsection{R\'{e}nyi entropies for WCFT}
The R\'{e}nyi entropies is quite straight forward to calculate after we have analysed the partition functions. It can be calculated by
\begin{align}
S_n=\frac{1}{1-n}\log \left(\frac{Z_{a|\bar{a}}(n\bar{\kappa}|n\kappa)}{Z_{a|\bar{a}}(\bar{\kappa}|\kappa)^n}\right).
\end{align}
Using (\ref{canonicalZ}) and the transformation rules of the partition function
\begin{align}\label{changeofpartitionfunctions}
Z_{a,\bar{a}}(\bar{\kappa}|\kappa)=e^{{k \bar{a}\over2}(\bar{\kappa}-\frac{\kappa \bar{a}}{2a})}Z_{1,0}\left(\bar{\kappa}-\frac{\bar{a}\kappa}{a}|\frac{\kappa}{a}\right)\,,
\end{align}
we find
\begin{align}
S_n=-i P_0^{vac} \left(\textit{l}_T+\frac{\bar{\beta}-\alpha}{\beta}\textit{l}_X\right)+\left(-i \frac{\alpha}{\pi}P_0^{vac}-\frac{2(n+1) L_0^{vac} }{n}\right) \log\left(\frac{\beta}{\pi\epsilon}\sinh\frac{\textit{l}_X \pi}{\beta}\right)\,.\label{renyi}
\end{align}

\subsection{Partition function and R\'{e}nyi entropy calculated on the gravity side}
Following the spirit of \cite{Hung}, in this subsection we calculate the R\'{e}nyi entropy holographically on the gravity side. As we have mentioned previously, the warping parameter does not affect the thermal properties of these WAdS spacetimes.  For simplicity we set $\lambda=0$, $k=-\frac{\ell}{G}$ and consider Einstein gravity with a negative cosmological constant in 3 dimensions.
Under the transformation (\ref{cthatx}), AdS$_{(\hat{u},\hat{v},\hat{r})}$ (\ref{wadshatu}) can be written in the $(\hat{x},\hat{t})$ coordinates
\begin{equation}\label{wbshxht}
ds^2=\ell^2\left(d\hat{t}^2+2\frac{r+T_{\hat{v}}^2}{T_{\hat{v}}}d\hat{t}d\hat{x}+(2r+T_{\hat{v}}^2+T_{\hat{u}}^2)d\hat{x}^2+\frac{dr^2}{4(r^2-T_{\hat{v}}^2T_{\hat{u}}^2)}\right)\,.
\end{equation}
The variation of conserved charges $\delta Q_{\xi}$ associate with the Killing vector $\xi^{\mu}$ between two backgrounds with metric $g+\delta g$ and $g$ is given by,
\begin{equation}
\delta Q_{\xi}=\int_{\partial\Sigma}\frac{1}{2}\epsilon_{\mu\nu\rho}K_{\xi}^{\mu\nu}(\delta g, g) dx^{\rho}\,,
\end{equation}
where $\epsilon_{\mu\nu\rho}$ is the volume form of metric (\ref{wbshxht}), and \begin{equation}
K_{\xi}^{\mu\nu}=\frac{1}{8\pi G}\left(\xi^{\nu}\nabla^{\mu}h-\xi^{\nu}\nabla_{\sigma}h^{\mu\sigma}+\xi_{\sigma}\nabla^{\nu}h^{\mu\sigma}+\frac{1}{2}h\nabla^{\nu}\xi^{\mu}-h^{\rho\nu}\nabla_{\rho}\xi^{\mu}\right)\,,
\end{equation}
where $h_{\mu\nu}=\delta g_{\mu \nu}\,, h^{\mu\nu}=g^{\mu\rho}g^{\nu\sigma}h_{\rho\sigma}$, $h=g^{\mu\nu}h_{\mu\nu}$, and $\nabla_{\mu}$ is the covariant derivative compatible with $g_{\mu\nu}$. Now, it is strightforward to calculate the variation of $L_0$ and $P_0$, which are charges associated with $\partial/\partial\hat{x}$ and $\partial/\partial\hat{t}$, and after a trivially integration we get
\begin{align}
L_0=\frac{\ell}{4G}(T_{\hat{u}}^2-T_{\hat{v}}^2)\,,\qquad P_0=-\frac{\ell}{2G}T_{\hat{v}}\,.
\end{align}
The generator of the horizon is taken to be,
\begin{equation}
\xi_H=\partial_{\hat{x}}+\Omega_H\partial_{\hat{t}}\,,
\end{equation}
which indicates the angular potential $\Omega_H$ is given by
\begin{align}
\Omega_H=-(T_{\hat{u}}+T_{\hat{v}})\,.
\end{align}
Also we can calculate the Hawking temperature and Bekenstein-Hawking entropy
\begin{equation}\label{TO}
T_H=\frac{T_{\hat{u}}}{\pi}\,,\qquad S=\frac{\pi\ell}{2G}(T_{\hat{v}}+T_{\hat{u}})\,.
\end{equation}
With all these physical quantities calculated, we find that the first law of thermodynamics
$
T_H\delta S=\delta L_0+\Omega_H\delta P_0
$
is satisfied.
The partition function of the gravity theory with angular momentum takes the form
\begin{equation}
Z(T_H, \Omega_H)=\exp\left(-\frac{1}{T_H}\Psi(T_H, \Omega_H)\right)\,,
\end{equation}
where $\Psi(T_H, \Omega_H)$ is the grand potential which is given by
\begin{align}
\Psi=L_0-T_HS+\Omega_HP_0=-\frac{\ell}{4G}\left(T_{\hat{u}}^2-T_{\hat{v}}^2\right)\,.
\end{align}
Thus the gravity partition function is given by
\begin{equation}\label{GP}
Z(T_{\hat{u}}, T_{\hat{v}})=\exp\left(\frac{\pi\ell}{4G}\frac{T_{\hat{u}}^2-T_{\hat{v}}^2}{T_{\hat{u}}}\right)\,.
\end{equation}
The R\'{e}nyi entropy is defined by
\begin{equation}
S_n=\frac{1}{1-n}\log\mathrm{Tr}(\rho^n)\,,
\end{equation}
where $\rho$ is the normalized density matrix
\begin{equation}
\rho=\frac{\exp\left[-\frac{1}{T_H}(\hat{L}_0+\Omega_H\hat{P}_0)\right]}{Z(T_H, \Omega_H)}\,.
\end{equation}
It can be shown that
\begin{eqnarray}
S_n^{bk}&=&\frac{1}{1-n}\log\left[\frac{Z(T_H/n, \Omega_H)}{Z(T_H, \Omega_H)^n}\right]\nonumber\\
&=&\frac{\pi\ell}{2G}(T_{\hat{v}}+T_{\hat{u}})\\
&=&S_{HEE}
\end{eqnarray}
where in the last line we have used the transformation between  $(\hu,\hv) $ and $(U,V)$
to rewrite $S_n^{bk} $ as the proposed holographic entanglement entropy (\ref{BHentropy}).

\subsection{Matching}
Using the vacuum values of the charges (\ref{P0L0vac}), we find that the R\'{e}nyi entropy is independent of $n$,
and is just the same as entanglement entropy
\be
S^{bn}_n=S_{EE}\,.
\ee
As was discussed in the previous section, the bulk and boundary calculations of entanglement entropy agree. Therefore,  the bulk and boundary calculations for the R\'{e}nyi entropy also agree,
\be  S_n^{bh}=S_{HEE}=S_{EE}=S_n^{bn}\,.\ee
Another consistent check is for the partition functions.
The torus of WCFT$_{(\hat x, \hat t)}$ is parametrized by
\begin{equation}\label{kbhkh}
\hat{\kappa}=\frac{\pi}{T_{\hat{u}}}\,,~~~~\hat{\bar{\kappa}}=\pi \left(1+\frac{T_{\hat{v}}}{T_{\hat{u}}}\right).
\end{equation}
Substituting   (\ref{kbhkh}) and  the vacuum charges (\ref{P0L0vac}) into (\ref{canonicalZ}), it is straightforward to show that,
\begin{equation}
Z_{1,0}(\hat{\bar{\kappa}}|\hat{\kappa})=\exp\left(\frac{\pi\ell}{4G}\frac{T_{\hat{u}}^2-T_{\hat{v}}^2}{T_{\hat{u}}}\right)\,,
\end{equation}
which is identical to the gravity partition function (\ref{GP}).

Therefore for the holographic WCFTs we considered, namely WCFTs dual to Einstein gravity (with matter) on (W)AdS$_3$ spacetimes with Dirichlet-Neumann boundary conditions, we find that the R\'{e}nyi entropy equals to entanglement entropy. This is due to the fact that for such theories $L_0^{vac}=0$. For a general WCFT, $L_0^{vac}$ is not necessarily zero, and the R\'{e}nyi entropy (\ref{renyi}) will in general depend on $n$.

 \section*{Acknowledgement}
We thank H.~Casini, A.~Castro, G.~Comp\`{e}re, T.~Hartman, D.~Hofman, H.~Jiang, R.~Miao, A.~Strominger and J.~Wu for helpful discussions. This work was supported in part by start-up funding 543310007 from Tsinghua University. W.S. is also supported by the National Thousand-Young-Talents Program of China.

\appendix
\section{The general quotient}\label{appendix1}
Here we give the general quotient of WAdS$_{(U,V,\rho)}$ (\ref{wbsT1T0}) with arbitrary parameters $a_0,a_+,a_-,a_L$. The corresponding coordinate transformation is given by
\begin{align}
u=&\frac{\coth ^{-1}\left(\frac{\sqrt{a_0^2-4 a_- a_+} \rho}{a_0 \rho-a_+ (2 \rho U+1)}\right)+\coth ^{-1}\left(\frac{\sqrt{a_0^2-4 a_- a_+} \rho}{-2 a_+ \rho U+a_0 \rho+a_+}\right)}{\sqrt{a_0^2-4 a_- a_+}}\,,
\cr
v=&V+\frac{1}{4} \log \left(\frac{4 a_- \rho^2+(2 \rho U+1) \left(a_+ (2 \rho U+1)-2 a_0 \rho\right)}{4 a_- \rho^2+(2 \rho U-1) \left(a_+ (2 \rho U-1)-2 a_0 \rho\right)}\right)
\cr
&~~~~+\frac{a_L \left(\coth ^{-1}\left(\frac{\sqrt{a_0^2-4 a_- a_+} \rho}{2 a_+ \rho U-a_0 \rho+a_+}\right)-\coth ^{-1}\left(\frac{\sqrt{a_0^2-4 a_- a_+} \rho}{-2 a_+ \rho U+a_0 \rho+a_+}\right)\right)}{\sqrt{a_0^2-4 a_- a_+}}\,,
\cr
r=&\frac{ \left(4 \rho \left(a_L+\rho U \left(a_+ U-a_0\right)+a_- \rho\right)-a_+\right)}{4 \rho}\,.
\end{align}
The resulting metric of WAdS$_{(u,v,r)}$ is
\begin{align}\label{wadsapp}
{ds^2\over\ell^2}=&\big(r ^2-\left(\lambda ^2+1\right) \left(r -r _-\right) \left(r -r _+\right)\big)\,du^2+ dv^2+2 r  \,du dv
+\frac{1}{4}  \frac{\left(\lambda ^2+1\right) }{(r -r_-) (r -r_+)}dr ^2\,,
\cr
r_-=& -\frac{1}{2}  \left(\sqrt{a_0^2-4 a_- a_+}-2 a_L\right)\,,\qquad r_+= \frac{1}{2}  \left(\sqrt{a_0^2-4 a_- a_+}+2 a_L\right)\,.
\end{align}
We see that the two horizon radius of WAdS$_{(u,v,r)}$ are  controlled by the way we do quotient. The metric can be transformed into the form of (\ref{blackstring}) with its two temperatures determined by $r_+$ and $r_-$. Under the choice of (\ref{ala0a+a-}), (\ref{wadsapp}) is already in the form of (\ref{blackstring}).

The asymptotic behavior of this coordinate transformation
\begin{align}
u=&\frac{2 \tanh ^{-1}\left(\frac{a_0-2 a_+ U}{\sqrt{a_0^2-4 a_- a_+}}\right)}{\sqrt{a_0^2-4 a_- a_+}}+\mathcal{O}\left(\frac{1}{\rho^2}\right)\,,
\cr
v=&V-\frac{2 a_L \tanh ^{-1}\left(\frac{a_0-2 a_+ U}{\sqrt{a_0^2-4 a_- a_+}}\right)}{\sqrt{a_0^2-4 a_- a_+}}+\mathcal{O}\left(\frac{1}{\rho}\right)\,.
\end{align}
indicate that the boundary of WAdS$_{(u,v,r)}$ covers a strip parallel to the $U$ axis on the boundary of WAdS$_{(U,V,\rho)}$, with $a_0/(2a_+)$ controlling the center position of the strip, while $l_U=-\sqrt{a_0^2-4 a_- a_+}/a_+$ controlling the width of the strip.

Consider the regulated interval (\ref{intervalUV'1}) in the main text, its image in terms of the $(u,v)$ coordinates is just (\ref{intervaltutv}) with
\begin{align}
\Delta u=&2\frac{\ell^2\log \left(\frac{l_U}{\epsilon}\right)}{r_+-r_-}\,,
\cr
\Delta v=&l_V-2 a_L \frac{\ell^2\log \left(\frac{l_U}{\epsilon}\right)}{r_+-r_-}\,.
\end{align}
As in the main text, we identify the end points and get a spatial circle $( u , v )\sim( u +\Delta  u , v +\Delta  v )$.  We integrate along the spatial circle at $r=r_+$, then get the thermal entropy for WAdS$_{(u,v,r)}$
\begin{align}\label{bhentropy}
S_{bh}&=\frac{1}{4 G} \int_{0}^{1} \sqrt{g_{uu}\Delta u^2+2 g_{uv}\Delta u\Delta v+g_{vv}\Delta v^2}d\tau
\cr
&=\frac{1}{4G}\frac{\left(\ell^2 \Delta v+r_+ \Delta u\right)}{\ell}
\cr
&=\frac{\ell}{4G}\log \frac{l_U}{\epsilon}+\frac{\ell}{4G}l_V\,.
\end{align}
It is then obvious that, among all the four quantities controlled by the way we do quotient, the Bekenstein-Hawking entropy for WAdS$_{(u,v,r)}$ only depend on the width of the strip. This justifies our choice in the main text.

\section{Rindler method in AdS$_3$ revisited}\label{appendix2}
In this Appendix, we revisit Rindler method in AdS$_3$ with our new strategy by doing quotient.  In Appendix \ref{A2.1}, we consider AdS$_3$ with Brown-Henneaux boundary conditions and give a derivation for the HRT formula \cite{HRT} in this context (AdS$_3$/CFT$_2$){\footnote{Recently a more general derivation for the HRT formula is given in \cite{Dong:2016hjy}.}. This generalize the derivation of \cite{CHM} to the covariant version. Then in Appendix \ref{A2.2} we consider AdS$_3$ with CSS boundary conditions, and find the holographic entanglement entropy agrees with the entanglement entropy of a WCFT.

According to the RT (or HRT) \cite{RT1,RT2,HRT} formula, the holographic entanglement entropy is proportional to a co-dimension two extremal surface, which is determined by the bulk metric. This indicates that the holographic entanglement entropy should be independent of the asymptotic boundary conditions. Here we consider the same metric with different boundary conditions and show explicitly how the choice of boundary conditions will change the holographic entanglement entropy for Poincar\'{e} AdS$_3$. This suggests that at least caution should be taken in applying RT (or HRT) proposal for holography beyond AdS/CFT.

Generalization to general temperatures can be done by considering the further mapping from BTZ black holes to Poincar\'{e} AdS$_3$ \cite{Carlip}, and then follow the steps of Sec. \ref{arbitrary temperatures} in the main text, or Sec. 5.5 of \cite{HRT}.
\subsection{Brown-Henneaux boundary conditions}\label{A2.1}
For simplicity we consider the Poincar\'{e} AdS$_3$ spacetime
\begin{align}\label{adsT0}
ds^2= \ell^2\Big(\frac{ d \rho^2}{4  \rho^2}+2  \rho d U d V\Big)\,.
\end{align}
Under the Brown-Henneaux boundary conditions, Poincar\'{e} AdS$_{3}$ is conjectured to be dual to the vacuum state of a CFT$_{2}$. The following six Killing vectors
\begin{align}\label{sixkv}
J_-=  \partial _U, \quad J_0=  U \p_{U}-\rho  \p_{\rho }, \quad J_+=  U^2\p_{U}-\frac{1}{2\rho }\p_{V}-2 \rho  U \p_{\rho },
\cr
\tilde{J}_-=  \partial _V, \quad \tilde{J}_0=  V\p_{V}-\rho \p_{\rho }, \quad \tilde{J}_+=  V^2\p_{V}-\frac{1}{2\rho }\p_{U}-2 \rho  V\p_{\rho }.
\end{align}
are also asymptotic Killing vectors.
The normalization are chosen to satisfy the standard $SL(2,R)\times SL(2,R)$ algebra
\begin{align}
&[J_-,J_+]=2J_0,\quad [J_0,J_\pm]=\pm J_{\pm},
\cr
&[\tilde{J}_-,\tilde{J}_+]=2\tilde{J}_0,\quad [\tilde{J}_0,\tilde{J}_\pm]=\pm \tilde{J}_{\pm}\,.
\end{align}

Now we do a coordinate transformation to obtain a new coordinate system with two explicit $U(1)$ symmetries. Define  \begin{align}
&J=a_0 J_0+a_+ J_++a_- J_-\,,
\cr
&\tilde{J}=\tilde{a}_0 \tilde{J}_0+\tilde{a}_+ \tilde{J}_++\tilde{a}_- \tilde{J}_-\,,
\end{align}
where
 $a_0,a_+,a_-,\tilde{a}_0,\tilde{a}_+,\tilde{a}_-$ are arbitrary constants.
Then we define new coordinates ($ u , v , r $) such that there are two explicit Killing vectors
\be \p_ u = J, \quad \p_ v =\tilde{J}\,. \ee
As a result the new metric in $(u,v,r)$ should only depend on $ r $.
Following the strategy of  Sec. \ref{4.1.1}, we can do the quotient on Poincar\'{e} AdS$_3$.  We find that the quotient gives a BTZ black string with its boundary covers the causal development of an interval. And the six parameters $a_0,a_+,a_-,\tilde{a}_0,\tilde{a}_+,\tilde{a}_-$ together control six quantities. Among which four quantities characterize the position and extension of an interval (or its causal development), and the rest two describe the two temperatures of BTZ$_{(u,v,r)}$. The regularized thermal entropy of BTZ$_{(u,v,r)}$ in fact only depends on the extension of the interval.

We choose the six parameters
\begin{align}
&a_0=0, \quad a_+=-\frac{2}{l_U}, \quad a_-=\frac{l_U}{2} \,,
\cr
&\tilde{a}_0=0, \quad \tilde{a}_+=-4 l_V{}^{-2}, \quad \tilde{ a}_-=1.
 \end{align}
 As we will see later, this choice is convenient for comparison with the CSS boundary conditions.
Similar to the steps in section \ref{secgravitystory}, we find the following coordinate transformation
\begin{align}\label{ctUuads}
 u =&\frac{1}{4}\log\Big[\frac{(1+ \rho(2 U+l_U) V)^2-  \rho^2 l_V^2(l_U/2+U)^2}{(1+ \rho(2U-l_U) V)^2-  \rho^2 l_V^2(l_U/2-U)^2}\Big]\,,
\cr
 v =&\frac{l_V}{8}  \log \Big[\frac{\left(\rho  \left(l_U-2 U\right) \left(l_V+2 V\right)-2\right) \left(\rho  \left(l_U+2 U\right) \left(l_V+2 V\right)+2\right)}{\left(\rho  \left(l_U-2 U\right) \left(l_V-2 V\right)+2\right) \left(\rho  \left(l_U+2 U\right) \left(l_V-2 V\right)-2\right)}\Big]\,,
\cr
 r =&\frac{l_U \rho}{2}(1- \frac{4 V^2}{l_V ^2})-\frac{2 \rho U^2}{l_U}+\frac{2(1+ 2\rho U V)^2}{ l_U l_V ^2 \rho}\,,
\end{align}
under which we get the metric of BTZ$_{(u,v,r)}$
\begin{align}\label{btztutv}
ds^2&=\ell^2\left(\,d u ^2+2  r   \,d u  d v + \frac{4}{l_V^2}d v ^2 + \frac{d r  ^2 }{4 ( r ^2 -4/l_V^2)}\right)\,.
\end{align}

The asymptotic behavior of the coordinate transformations is given by
\begin{align}
u=&\text{ArcTanh}\frac{2 U}{l_U}+\mathcal{O}\left(\frac{1}{r}\right)\,,
\cr
v=&\frac{l_V}{2} \text{ArcTanh}\left(\frac{2 V}{l_V}\right)+\mathcal{O}\left(\frac{1}{r}\right)\,,
\end{align}
which indicates the boundary of BTZ$_{(u,v,r)}$ (\ref{btztutv}) covers the causal development of an interval
\begin{align}
\{(U,\, V)|\,-\frac{l_U}{2}<U<\frac{l_U}{2} \,,\quad -\frac{l_V}{2} <V<\frac{l_V}{2}\} \,.
\end{align}
We can introduce two infinitesimal parameters $\epsilon_1$ and $\epsilon_2$ and regulate the interval as
\begin{align}\label{rUV}
\{(U,\,V)|\, U={(l_U-2\epsilon_1) }(-{1\over2}+\tau) ,\quad V={(l_V-2\epsilon_2)}(-{1\over2}+\tau),\quad \tau\in[0,1]\}\,.
 \end{align}
The horizon of BTZ$_{(u,v,r)}$ can be depicted by
\begin{align}
&\{(u,\,v)|\, u={\Delta u }(-{1\over2}+\tau) ,\quad v={\Delta v}(-{1\over2}+\tau),\quad r=\frac{2}{l_V},\quad \tau\in[0,1]\}\,,
\cr
&\Delta u=\log\frac{l_U}{\epsilon_1}\,,
\qquad
\Delta v=\frac{l_V}{2}\log\frac{ l_V}{\epsilon_2}\,.
\end{align}

Since $\Delta u$ and $\Delta v$ are infinite, as in the main text we identify the end points and get a spatial circle $(u,v)\sim (u+\Delta u,v+\Delta v)$, thus the Bekenstein-Hawking entropy of BTZ$_{(u,v,r)}$ (\ref{btztutv}) is given by
\begin{align}
S_{thermal}=\frac{\ell}{4 G}\log\frac{l_U l_V}{\epsilon_1\epsilon_2}\,.
\end{align}

Furthermore, to recover the entangling surface in the original vacuum AdS$_3$ spacetime, we can look at the inverse coordinate transformation
\begin{align}\label{ctAdS1}
&U = \frac{l_U}{2}\frac{  \sqrt{r ^2-4 l_V^{-2}   } \sinh (u +2 v/l_V   )-\left(r- 2 l_V^{-1}    \right) \sinh (u - 2 v/l_V   )}{\sqrt{r^2-4 l_V^{-2}   } \cosh (u +2 v/l_V   )-\left(r- 2 l_V^{-1}    \right) \cosh (u -2 v/l_V   )}\,,
\cr
&V = \frac{l_V}{2}\frac{ \sqrt{r ^2-4 l_V^{-2}   } \sinh (u +2 v/l_V   )+\left(r -2 l_V^{-1}   \right) \sinh (u -2 v/l_V   )}{\sqrt{ r ^2-4 l_V^{-2}   } \cosh (u +2 v/l_V   )-  \left(r- 2 l_V^{-1}    \right) \cosh (u -2 v/l_V   )}\,,
\cr
&\rho=\frac{\left(\sqrt{r ^2-4 l_V^{-2}   } \cosh (u +2 v/l_V   )-\left(r- 2 l_V^{-1}    \right) \cosh (u -2 v/l_V   )\right)^2}{2 l_U  \left(r-2 l_V^{-1}   \right)}\,.
\end{align}
The image of the BTZ black hole horizon in AdS$_3$ is given by
\begin{align}\label{geodesic}
U=&\frac{l_U}{2}\tanh \tilde{\tau}\,,
\qquad
 V=\frac{l_V}{2}\tanh \tilde{\tau}\,,
\qquad
\rho=\frac{2\cosh^2 \tilde{\tau}}{l_U l_V}\,,
\end{align}
where $\tilde{\tau}$ is defined by
\begin{align}
\tilde{\tau}=&u +2 v/l_V=\left(\tau -\frac{1}{2}\right)\log\frac{l_U l_V }{\epsilon_1\epsilon_2}\,.
\end{align}
One can check that (\ref{geodesic}) is just the geodesic ending on $(-\frac{l_U}{2},-\frac{l_V}{2})$ and $(\frac{l_U}{2},\frac{l_V}{2})$ on the AdS$_{3}$ boundary.

Note that when $\tau=0,1$, the geodesic goes to the two end points on the boundary with the same cutoff
\begin{align}
 \rho_{max}=\frac{1}{2\epsilon_1\epsilon_2}+\frac{1}{l_Ul_V}+\mathcal{O}\left(\epsilon^2\right).
 \end{align}
 This indicates that the cutoff of the theory should be $\epsilon^2=\epsilon_1\epsilon_2$, and hence
 \begin{align}\label{ads1}
 S_{thermal}=\frac{\ell}{4 G}\log\frac{l_U l_V}{\epsilon^2}
 \end{align}
which agrees with the HRT formula.

\subsection{CSS (Dirichlet-Neumann) boundary conditions}\label{A2.2}
For comparison, in this subsection we impose the CSS boundary conditions on AdS$_3$, hence the dual field theory is conjectured to be a WCFT \cite{NBC} (see Sec. \ref{adswcft}).
To compare with the previous subsection, we also consider Poincar\'{e} AdS$_3$. All the calculations are the same as in section \ref{arbitrary temperatures} with $T_V=T_U=\lambda=0$. Below we will list the main differences between AdS/CFT and AdS/WCFT.

The first difference is the choice of Killing vectors. Under the CSS boundary conditions, $\tilde{J}_+$ and $\tilde{J}_0$, though still isometry generators, are no longer asymptotic Killing vectors. Hence we need to set $\tilde{a}_0=\tilde{a}_+=0$ to do the quotient. We have four parameters to do quotient under CSS, while six under Brown-Henneaux. As in section \ref{sectu0tv1}, we choose
\begin{align}\label{as2}
&a_0=0, \quad a_+=-\frac{2}{l_U}, \quad a_-=\frac{l_U}{2}, \quad \tilde{ a}_-=1.
 \end{align}
Then the quotient is just the case in Sec. \ref{arbitrary temperatures} with $\lambda=0,~T_U=T_V=0$, and we get a BTZ$_{(u,v,r)}$.

The second difference is that $l_V$ in this case is a free parameter rather than controlled by the way we do quotient. Under CSS, we can chose $l_U$ and $l_V$ to be the same as in Appendix \ref{A2.1}, while the thermal entropy of BTZ$_{(u,v,r)}$ is given by (\ref{BHentropy}) with $T_V=T_U=0$,
\begin{align}\label{ads2}
S_{thermal}=\frac{\ell}{4 G}\log\frac{ l_U}{\epsilon}\,.
\end{align}

The disagreement between (\ref{ads1}) and (\ref{ads2}), which give the holographic entanglement entropy for the same interval while under different boundary conditions, is of course a result of the changing of boundary conditions.

\section{The quotient: from WAdS$_{(T,X,\rho)}$ to WAdS$_{(t,x,r)}$}
As stated in section (\ref{5.1}), a state-dependent coordinate transformation (\ref{ctXU}) is needed once we require that the bulk asymptotic analysis of WAdS yields a canonical WCFT algebra (\ref{wcftalg}). Let WAdS$_{(T,X,\rho)}$ denotes the WAdS in the $\{T, X, \rho\}$ coordinate system. In this subsection, we redo the $SL(2,R)\times U(1)$ quotient on that of WAdS$_{(T,X,\rho)}$ as a consistent check, and the resulting quotient space is the an analog of hyperbolic black hole WAdS$_{(t,x,r)}$ where the subscript denotes the coordinate system.

We are considering a bulk WAdS with $T_U=0, T_V=1$. After the coordinate transformation (\ref{ctXU}), the spacetime metric becomes,
\begin{equation}\label{ds2TX}
ds^2=\ell^2\left(-\lambda^2\rho^2dX^2+2\rho dX\left(\sqrt{-\frac{Gk}{\ell}}dT+dX\right)+\left(\sqrt{-\frac{Gk}{\ell}}dT+dX\right)^2+\frac{1+\lambda^2}{4\rho^2}d\rho^2\right)\,.
\end{equation}
The Killing vectors are still those of $J_{\pm}, J_0$, and $J_L$ in Eqs. (\ref{KVs}) by dong a coordinate transformation (\ref{ctXU}). These Killing vectors form a $SL(2,R)\times U(1)$ algebra as spacetime (\ref{ds2TX}) symmetry required. By doing quotient, we are trying to find a new coordinate system $\{t, x, r\}$ such that $\partial_t$ and $\partial_x$ are explicit commuting Killing vectors. Without lose of generality, these two Killing coordinates can be defined as,
\begin{equation}
\partial_t=J_L,~~~~\partial_x=J\,,
\end{equation}
where $J=a_LJ_L+a_+J_++a_0J_0+a_-J_-$ and $a_L$, $a_+$, $a_0$, and $a_-$ are four arbitrary constants. The new radial coordinate can be chosen as,
\begin{equation}
r\equiv{J\cdot J_L\over \ell^2}\,.
\end{equation}
According to the same procedure as in section (\ref{4.1.1}), the boundary coordinates $(t, x)$ cover a strip of $(T, X)$ with width $l_X=-\sqrt{a_0^2-4a_+a_-}/a_+$. So the four parameters $a_L$, $a_+$, $a_0$, and $a_-$ determine the interval on the boundary with redundancy. For simplicity, these four parameters can be chosen as
\begin{equation}
a_L=a_0=0,~~~~a_+=-\frac{2}{l_X},~~~~a_-=\frac{l_X}{2}\,,
\end{equation}
without changing the result. In such parametrization, the coordinate transformations take the form,
\begin{eqnarray}
t&=&T+\sqrt{-\frac{\ell}{Gk}}X+\frac{1}{4}\sqrt{-\frac{\ell}{Gk}}\log\left(\frac{(1+2X\rho)^2-l_X^2\rho^2}{(1-2X\rho)^2-l_X^2\rho^2}\right)\,,\\
x&=&\frac{1}{4}\log\left(\frac{(l_X+2X)^2\rho^2-1}{(l_X-2X)^2\rho^2-1}\right)\,,\\
r&=&\frac{1+(l_X^2-4X^2)\rho^2}{2l_X\rho}\,.
\end{eqnarray}
and the resulting metric in coordinate system $\{t, x, r\}$ becomes,
\begin{equation}\label{hpbhtx}
ds^2=\ell^2\left((1+\lambda^2-\lambda^2r^2)dx^2+2rdx\left(\sqrt{-\frac{Gk}{\ell}}dt+dx\right)+\left(\sqrt{-\frac{Gk}{\ell}}dt+dx\right)^2+\frac{1+\lambda^2}{4(r^2-1)}dr^2\right)\,.
\end{equation}
This is just the black string metric with $T_u=1$ and $T_v=1$. The coordinate transformations on the boundary are
\begin{eqnarray}\label{TXtxBundary}
t&=&T+\sqrt{-\frac{\ell}{Gk}}X+\mathcal{O}\left(\frac{1}{\rho}\right)\,,\\
x&=&\frac{1}{2}\log\left(\frac{l_X+2X}{l_X-2X}\right)+\mathcal{O}\left(\frac{1}{\rho}\right)^2\,.
\end{eqnarray}
Let us consider an interval on the boundary $\rho\to\infty$,
\begin{equation}
\mathcal{A}:~~\{(T,\,X)|\, T=l_T\left(-\frac{1}{2}+\tau\right),\quad X=(l_X-2\epsilon)\left(-\frac{1}{2}+\tau\right),\quad \tau\in[0,1]\}\,,
\end{equation}
where $\epsilon$ is a small cutoff. The above interval covers the $(t, x)$ coordinates with range,
\begin{equation}
\{(t,\,x)|\,t=\Delta t\left(-\frac{1}{2}+\tau\right),\quad x=\Delta x\left(-\frac{1}{2}+\tau\right),\quad \tau\in[0,1]\}\,,
\end{equation}
where
\begin{equation}
\Delta t=l_T+\sqrt{-\frac{\ell}{Gk}}l_X,~~~~\Delta x=\log\frac{l_X}{\epsilon}\,.
\end{equation}
The outer horizon area of the black string (\ref{hpbhtx}) divided by $4G$ measures the thermal entropy of the black string,
\begin{eqnarray}
S_{HEE}&=&\frac{1}{4G}\int_0^1\sqrt{g_{tt}\Delta t^2+2g_{tx}\Delta t\Delta x+g_{xx}\Delta x^2}d\lambda\nonumber\\
&=&\frac{\ell}{4G}\left(\sqrt{-\frac{Gk}{\ell}}l_T+l_X+\log\frac{l_X}{\epsilon}\right)\,.
\end{eqnarray}
This thermal entropy matches that of the bulk calculation of the entanglement entropy (\ref{BHentropy01}) with $T_U=0, T_V=1$.


\begin{thebibliography}{}
\bibitem{SWX}
  W.~Song, Q.~Wen and J.~Xu,
  ``Generalized Gravitational Entropy for Warped Anti-de Sitter Space,''
  Phys.\ Rev.\ Lett.\  {\bf 117}, no. 1, 011602 (2016)
  [arXiv:1601.02634 [hep-th]].



\bibitem{ADSCFT1}
  J.~M.~Maldacena,
  ``The Large N limit of superconformal field theories and supergravity,''
  Int.\ J.\ Theor.\ Phys.\  {\bf 38}, 1113 (1999).
  [Adv.\ Theor.\ Math.\ Phys.\  {\bf 2}, 231 (1998).]
  [arXiv:hep-th/9711200 [hep-th]];

  S.~S.~Gubser, I.~R.~Klebanov and A.~M.~Polyakov,
 ``Gauge theory correlators from noncritical string theory,''
  Phys.\ Lett.\ B {\bf 428}, 105 (1998).
 [arXiv:hep-th/9802109 [hep-th]];

  E.~Witten,
 ``Anti-de Sitter space and holography,''
  Adv.\ Theor.\ Math.\ Phys.\  {\bf 2}, 253 (1998).
 [arXiv:hep-th/9802150 [hep-th]].

\bibitem{dscft}
  A.~Strominger,
  ``The dS / CFT correspondence,''
  JHEP {\bf 0110}, 034 (2001)
  [hep-th/0106113];

  D.~Anninos, T.~Hartman and A.~Strominger,
  ``Higher Spin Realization of the dS/CFT Correspondence,''
  arXiv:1108.5735 [hep-th].


\bibitem{KerrCFT1}
  M.~Guica, T.~Hartman, W.~Song and A.~Strominger,
  ``The Kerr/CFT Correspondence,''
  Phys.\ Rev.\ D {\bf 80}, 124008 (2009)
  [arXiv:0809.4266 [hep-th]];

  I.~Bredberg, T.~Hartman, W.~Song and A.~Strominger,
  ``Black Hole Superradiance From Kerr/CFT,''
  JHEP {\bf 1004}, 019 (2010)
  [arXiv:0907.3477 [hep-th]];

  A.~Castro, A.~Maloney and A.~Strominger,
  ``Hidden Conformal Symmetry of the Kerr Black Hole,''
  Phys.\ Rev.\ D {\bf 82}, 024008 (2010)
  [arXiv:1004.0996 [hep-th]];

  I.~Bredberg, C.~Keeler, V.~Lysov and A.~Strominger,
 ``Cargese Lectures on the Kerr/CFT Correspondence,''
  Nucl.\ Phys.\ Proc.\ Suppl.\  {\bf 216}, 194 (2011).
 [arXiv:1103.2355 [hep-th]];

  G.~Compere,
  ``The Kerr/CFT correspondence and its extensions: a comprehensive review,''
  Living Rev.\ Rel.\  {\bf 15}, 11 (2012)
  [arXiv:1203.3561 [hep-th]].

\bibitem{warpedblackholes}
  D.~Anninos, W.~Li, M.~Padi, W.~Song and A.~Strominger,
  ``Warped AdS(3) Black Holes,''
  JHEP {\bf 0903}, 130 (2009)
  [arXiv:0807.3040 [hep-th]].

\bibitem{DHH}
  S.~Detournay, T.~Hartman and D.~M.~Hofman,
  ``Warped Conformal Field Theory,''
  Phys.\ Rev.\ D {\bf 86}, 124018 (2012)
  [arXiv:1210.0539 [hep-th]].




\bibitem{flat1}
  G.~Barnich and C.~Troessaert,
 ``Aspects of the BMS/CFT correspondence,''
  JHEP {\bf 1005}, 062 (2010).
 [arXiv:1001.1541 [hep-th]];

  A.~Bagchi,
  ``The BMS/GCA correspondence,''
  arXiv:1006.3354 [hep-th];

  A.~Bagchi,
  ``Correspondence between Asymptotically Flat Spacetimes and Nonrelativistic Conformal Field Theories,''
  Phys.\ Rev.\ Lett.\  {\bf 105}, 171601 (2010);

  A.~Bagchi and R.~Fareghbal,
  ``BMS/GCA Redux: Towards Flatspace Holography from Non-Relativistic Symmetries,''
  JHEP {\bf 1210}, 092 (2012)
  [arXiv:1203.5795 [hep-th]];

  A.~Bagchi, R.~Basu, D.~Grumiller and M.~Riegler,
  ``Entanglement entropy in Galilean conformal field theories and flat holography,''
  Phys.\ Rev.\ Lett.\  {\bf 114}, no. 11, 111602 (2015)
  [arXiv:1410.4089 [hep-th]];

  S.~M.~Hosseini and A.~Veliz-Osorio,
  ``Gravitational anomalies, entanglement entropy, and flat-space holography,''
  Phys.\ Rev.\ D {\bf 93}, no. 4, 046005 (2016).
  [arXiv:1507.06625 [hep-th]];

  S.~M.~Hosseini and A.~Veliz-Osorio,
  ``Entanglement and mutual information in two-dimensional nonrelativistic field theories,''
  Phys.\ Rev.\ D {\bf 93}, no. 2, 026010 (2016).
  [arXiv:1510.03876 [hep-th]].



\bibitem{BMS}
  H.~Bondi, M.~G.~J.~van der Burg and A.~W.~K.~Metzner,
  ``Gravitational waves in general relativity. 7. Waves from axisymmetric isolated systems,''
Proc.\ Roy.\ Soc.\ Lond.\ A {\bf 269}, 21 (1962);

  R.~K.~Sachs,
  ``Gravitational waves in general relativity. 8. Waves in asymptotically flat space-times,''
Proc.\ Roy.\ Soc.\ Lond.\ A {\bf 270}, 103 (1962);

  A.~Strominger,
  ``On BMS Invariance of Gravitational Scattering,''
  JHEP {\bf 1407}, 152 (2014)
  [arXiv:1312.2229 [hep-th]];

  S.~W.~Hawking, M.~J.~Perry and A.~Strominger,
  Phys.\ Rev.\ Lett.\  {\bf 116}, no. 23, 231301 (2016)
  [arXiv:1601.00921 [hep-th]].



\bibitem{lifshitz1}
  D.~T.~Son,
 ``Toward an AdS/cold atoms correspondence: A Geometric realization of the Schrodinger symmetry,''
  Phys.\ Rev.\ D {\bf 78}, 046003 (2008).
 [arXiv:0804.3972 [hep-th]];

  K.~Balasubramanian and J.~McGreevy,
 ``Gravity duals for non-relativistic CFTs,''
  Phys.\ Rev.\ Lett.\  {\bf 101}, 061601 (2008).
 [arXiv:0804.4053 [hep-th]];

  S.~Kachru, X.~Liu and M.~Mulligan,
 ``Gravity duals of Lifshitz-like fixed points,''
  Phys.\ Rev.\ D {\bf 78}, 106005 (2008).
 [arXiv:0808.1725 [hep-th]];

  M.~Taylor,
  ``Lifshitz holography,''
  Class.\ Quant.\ Grav.\  {\bf 33}, no. 3, 033001 (2016).
  [arXiv:1512.03554 [hep-th]].

\bibitem{Anninos08}
  D.~Anninos,
  ``Hopfing and Puffing Warped Anti-de Sitter Space,''
  JHEP {\bf 0909}, 075 (2009)
  [arXiv:0809.2433 [hep-th]].

\bibitem{CGR}
  G.~Comp\`{e}re, M.~Guica and M.~J.~Rodriguez,
 ``Two Virasoro symmetries in stringy warped AdS$_{3}$,''
  JHEP {\bf 1412}, 012 (2014).
 [arXiv:1407.7871 [hep-th]].


\bibitem{Compere-Detournay}
  G.~Comp\`{e}re and S.~Detournay,
 ``Boundary conditions for spacelike and timelike warped AdS$_3$ spaces in topologically massive gravity,''
  JHEP {\bf 0908}, 092 (2009).
 [arXiv:0906.1243 [hep-th]].

\bibitem{Cardyformula}
  J.~L.~Cardy,
  ``Operator Content of Two-Dimensional Conformally Invariant Theories,''
  Nucl.\ Phys.\ B {\bf 270}, 186 (1986).


\bibitem{Hofman-Strominger}
   D.~M.~Hofman and A.~Strominger,
 ``Chiral Scale and Conformal Invariance in 2D Quantum Field Theory,''
   Phys.\ Rev.\ Lett.\  {\bf 107}, 161601 (2011).
  [arXiv:1107.2917 [hep-th]].


\bibitem{ChiralLiouville}
  G.~Comp\`ere, W.~Song and A.~Strominger,
   ``Chiral Liouville Gravity,''
  JHEP {\bf 1305}, 154 (2013)
  [arXiv:1303.2660 [hep-th]].


\bibitem{Hofman:2014loa}
  D.~M.~Hofman and B.~Rollier,
  ``Warped Conformal Field Theory as Lower Spin Gravity,''
  Nucl.\ Phys.\ B {\bf 897}, 1 (2015)
  [arXiv:1411.0672 [hep-th]].


\bibitem{Castro-Hofman-Sarosi}
  A.~Castro, D.~M.~Hofman and G.~S\'arosi,
  ``Warped Weyl fermion partition functions,''
  JHEP {\bf 1511} (2015) 129
  [arXiv:1508.06302 [hep-th]].


\bibitem{NBC}
  G.~Comp\`{e}re, W.~Song and A.~Strominger,
 ``New Boundary Conditions for AdS3,''
  JHEP {\bf 1305}, 152 (2013).
 [arXiv:1303.2662 [hep-th]].


\bibitem{Guica-Strominger}
  M.~Guica and A.~Strominger,
  ``Microscopic Realization of the Kerr/CFT Correspondence,''
  JHEP {\bf 1102}, 010 (2011)
  [arXiv:1009.5039 [hep-th]].

\bibitem{ElShowk-Guica}
  S.~El-Showk and M.~Guica,
  ``Kerr/CFT, dipole theories and nonrelativistic CFTs,''
  JHEP {\bf 1212}, 009 (2012)
  [arXiv:1108.6091 [hep-th]].

\bibitem{Song-Strominger}
  W.~Song and A.~Strominger,
  ``Warped AdS3/Dipole-CFT Duality,''
  JHEP {\bf 1205}, 120 (2012)
  [arXiv:1109.0544 [hep-th]].

\bibitem{Detournay-Guica}
  S.~Detournay and M.~Guica,
 ``Stringy Schr$\ddot{\text{o}}$dinger truncations,''
  JHEP {\bf 1308}, 121 (2013).
 [arXiv:1212.6792 [hep-th]].

\bibitem{RT1}
  S.~Ryu and T.~Takayanagi,
 ``Holographic derivation of entanglement entropy from AdS/CFT,''
  Phys.\ Rev.\ Lett.\  {\bf 96}, 181602 (2006).
 [arXiv:hep-th/0603001 [hep-th]].

\bibitem{RT2}
  S.~Ryu and T.~Takayanagi,
 ``Aspects of Holographic Entanglement Entropy,''
  JHEP {\bf 0608}, 045 (2006).
 [arXiv:hep-th/0605073 [hep-th]].



\bibitem{CHM}
  H.~Casini, M.~Huerta and R.~C.~Myers,
   ``Towards a derivation of holographic entanglement entropy,''
  JHEP {\bf 1105}, 036 (2011)
  [arXiv:1102.0440 [hep-th]].

\bibitem{Hartman}
  T.~Hartman,
  ``Entanglement Entropy at Large Central Charge,''
  arXiv:1303.6955 [hep-th].

\bibitem{Faulkner}
  T.~Faulkner,
  ``The Entanglement Renyi Entropies of Disjoint Intervals in AdS/CFT,''
  arXiv:1303.7221 [hep-th].

\bibitem{LM}
  A.~Lewkowycz and J.~Maldacena,
 ``Generalized gravitational entropy,''
  JHEP {\bf 1308}, 090 (2013).
 [arXiv:1304.4926 [hep-th]].

\bibitem{HRT}
  V.~E.~Hubeny, M.~Rangamani and T.~Takayanagi,
 ``A Covariant holographic entanglement entropy proposal,''
  JHEP {\bf 0707}, 062 (2007).
 [arXiv:0705.0016 [hep-th]].


\bibitem{Anninos13}
  D.~Anninos, J.~Samani and E.~Shaghoulian,
  ``Warped Entanglement Entropy,''
  JHEP {\bf 1402}, 118 (2014)
  [arXiv:1309.2579 [hep-th]].

\bibitem{Basanisi:2016hsh}
  L.~Basanisi and S.~Chakrabortty,
  ``Holographic Entanglement Entropy in NMG,''
  JHEP {\bf 1609}, 144 (2016)
  [arXiv:1606.01920 [hep-th]].

\bibitem{CHI}
  A.~Castro, D.~M.~Hofman and N.~Iqbal,
  ``Entanglement Entropy in Warped Conformal Field Theories,''
  JHEP {\bf 1602}, 033 (2016)
  [arXiv:1511.00707 [hep-th]].


\bibitem{Song:2011sr}
  W.~Song and A.~Strominger,
 ``Warped AdS3/Dipole-CFT Duality,''
  JHEP {\bf 1205}, 120 (2012).
 [arXiv:1109.0544 [hep-th]].






\bibitem{Troessaert:2013fma}
  C.~Troessaert,
  ``Enhanced asymptotic symmetry algebra of $AdS_{3}$,''
  JHEP {\bf 1308}, 044 (2013)
  [arXiv:1303.3296 [hep-th]];

  S.~G.~Avery, R.~R.~Poojary and N.~V.~Suryanarayana,
  ``An sl(2,$\mathbb{R}$) current algebra from $AdS_3$ gravity,''
  JHEP {\bf 1401}, 144 (2014)
  [arXiv:1304.4252 [hep-th]];

  C.~Troessaert,
  ``Poisson Structure of the Boundary Gravitons in 3D Gravity with Negative $\Lambda$,''
  Class.\ Quant.\ Grav.\  {\bf 32}, no. 23, 235019 (2015)
  [arXiv:1507.01580 [gr-qc]];

  L.~Donnay, G.~Giribet, H.~A.~Gonzalez and M.~Pino,
  ``Supertranslations and Superrotations at the Black Hole Horizon,''
  Phys.\ Rev.\ Lett.\  {\bf 116}, no. 9, 091101 (2016)
  [arXiv:1511.08687 [hep-th]];

  H.~Afshar, S.~Detournay, D.~Grumiller and B.~Oblak,
  ``Near-Horizon Geometry and Warped Conformal Symmetry,''
  JHEP {\bf 1603}, 187 (2016)
  [arXiv:1512.08233 [hep-th]];

  H.~Afshar, S.~Detournay, D.~Grumiller, W.~Merbis, A.~Perez, D.~Tempo and R.~Troncoso,
  ``Soft Heisenberg hair on black holes in three dimensions,''
  Phys.\ Rev.\ D {\bf 93}, no. 10, 101503 (2016)
  [arXiv:1603.04824 [hep-th]];

  A.~P\'erez, D.~Tempo and R.~Troncoso,
  ``Boundary conditions for General Relativity on AdS$_{3}$ and the KdV hierarchy,''
  JHEP {\bf 1606}, 103 (2016)
  [arXiv:1605.04490 [hep-th]];

  D.~Grumiller and M.~Riegler,
  ``Most general AdS$_3$ boundary conditions,''
  arXiv:1608.01308 [hep-th].

\bibitem{TMG1}
  S.~Deser, R.~Jackiw and S.~Templeton,
  ``Topologically Massive Gauge Theories,''
  Annals Phys.\  {\bf 140}, 372 (1982)
  [Annals Phys.\  {\bf 281}, 409 (2000)]
  Erratum: [Annals Phys.\  {\bf 185}, 406 (1988)].

\bibitem{TMG2}
  S.~Deser, R.~Jackiw and S.~Templeton,
  ``Three-Dimensional Massive Gauge Theories,''
  Phys.\ Rev.\ Lett.\  {\bf 48}, 975 (1982).

\bibitem{NMG}
  E.~A.~Bergshoeff, O.~Hohm and P.~K.~Townsend,
  ``Massive Gravity in Three Dimensions,''
  Phys.\ Rev.\ Lett.\  {\bf 102}, 201301 (2009)
  [arXiv:0901.1766 [hep-th]].


\bibitem{Bena-Guica-Song}
  I.~Bena, M.~Guica and W.~Song,
  ``Un-twisting the NHEK with spectral flows,''
  JHEP {\bf 1303}, 028 (2013)
  [arXiv:1203.4227 [hep-th]].


\bibitem{Azeyanagi-Hofman-Song-Strominger}
  T.~Azeyanagi, D.~M.~Hofman, W.~Song and A.~Strominger,
  ``The Spectrum of Strings on Warped AdS$_3\times S^3$,''
  JHEP {\bf 1304}, 078 (2013)
  [arXiv:1207.5050 [hep-th]].



\bibitem{CHMbase1}
  P.~Candelas and J.~S.~Dowker,
  ``Field Theories On Conformally Related Space-times: Some Global Considerations,''
  Phys.\ Rev.\ D {\bf 19}, 2902 (1979).

\bibitem{CHMbase2}
  R.~C.~Myers and A.~Sinha,
  ``Seeing a c-theorem with holography,''
  Phys.\ Rev.\ D {\bf 82}, 046006 (2010)
  [arXiv:1006.1263 [hep-th]].

\bibitem{CHMbase3}
  H.~Casini and M.~Huerta,
  ``Entanglement entropy for the n-sphere,''
  Phys.\ Lett.\ B {\bf 694}, 167 (2011)
  [arXiv:1007.1813 [hep-th]].

\bibitem{CHMbase4}
  R.~C.~Myers and A.~Sinha,
  ``Holographic c-theorems in arbitrary dimensions,''
  JHEP {\bf 1101}, 125 (2011)
  [arXiv:1011.5819 [hep-th]].

\bibitem{Hartman-Keller}
  T.~Hartman, C.~A.~Keller and B.~Stoica,
  ``Universal Spectrum of 2d Conformal Field Theory in the Large c Limit,''
  JHEP {\bf 1409}, 118 (2014)
  [arXiv:1405.5137 [hep-th]].

\bibitem{Hung}
  L.~Y.~Hung, R.~C.~Myers, M.~Smolkin and A.~Yale,
  ``Holographic Calculations of Renyi Entropy,''
  JHEP {\bf 1112}, 047 (2011)
  [arXiv:1110.1084 [hep-th]].

\bibitem{BHTZ}
  M.~Banados, M.~Henneaux, C.~Teitelboim and J.~Zanelli,
  ``Geometry of the (2+1) black hole,''
  Phys.\ Rev.\ D {\bf 48}, 1506 (1993)
  Erratum: [Phys.\ Rev.\ D {\bf 88}, 069902 (2013)]
  [gr-qc/9302012].

\bibitem{Dong:2016hjy}
  X.~Dong, A.~Lewkowycz and M.~Rangamani,
  ``Deriving covariant holographic entanglement,''
  JHEP {\bf 1611}, 028 (2016)
  [arXiv:1607.07506 [hep-th]].

\bibitem{Carlip}
  S.~Carlip and C.~Teitelboim,
  ``Aspects of black hole quantum mechanics and thermodynamics in (2+1)-dimensions,''
  Phys.\ Rev.\ D {\bf 51}, 622 (1995)
  [gr-qc/9405070].


\end{thebibliography}
\end{document}